\documentclass[%
reprint,
superscriptaddress,
 amsmath,amssymb,
 aps,
pra,
]{revtex4-2}

\usepackage{graphicx}
\usepackage{dcolumn}
\usepackage{bm}
\usepackage{hyperref}
\usepackage{color} 
\usepackage{txfonts}
\usepackage[dvipsnames]{xcolor}
\usepackage{ragged2e}
\usepackage{amsmath}

\begin{document}

\title{Fast dynamical control of quantum droplets via Feshbach resonances}

\author{Jing Li}
\email{jli@ucc.ie}
 \affiliation{School of Physics and Technology, Nantong University, Nantong 226019, China}
 \affiliation{School of Physics, University College Cork, Cork T12 K8AF, Ireland}

\author{Thom\'as Fogarty}
\affiliation{Quantum Systems Unit, Okinawa Institute of Science and Technology Graduate University, Onna, Okinawa 904-0495, Japan}

\author{Thomas Busch}
\affiliation{Quantum Systems Unit, Okinawa Institute of Science and Technology Graduate University, Onna, Okinawa 904-0495, Japan}

\author{Andreas Ruschhaupt}
\affiliation{School of Physics, University College Cork, Cork T12 K8AF, Ireland}
\affiliation{Rinn Quantum, School of Physics, University College Cork}


\begin{abstract}
We propose a variational shortcut-to-adiabaticity scheme for the fast dynamical control of one-dimensional quantum droplets in ultracold Bose-Bose mixtures. The control is implemented through time-dependent mean-field (MF) and beyond-mean-field (BMF) nonlinearities, which can be adjusted by tuning intra- and interspecies scattering lengths near Feshbach resonances. Using a bright-droplet variational ansatz, together with inverse engineering, we derive time-dependent interaction protocols that connect prescribed initial and target self-bound states while suppressing residual excitations. We analyze three representative settings: simultaneous control of the MF and BMF terms, control with fixed BMF nonlinearity and time-dependent MF interaction, and control with fixed MF nonlinearity and time-dependent BMF interaction. For the constrained cases, additional variational degrees of freedom are introduced to optimize the protocols while respecting the imposed restrictions. Direct numerical simulations of the extended Gross-Pitaevskii equation confirm final fidelities exceeding $0.99$ on short time scales, and show that the optimised protocols are robust against finite calibration errors of the nonlinear control strength. Our results provide a route toward fast state-to-state manipulation of self-bound quantum fluids and may be useful for controlled matter-wave engineering in ultracold mixtures.
\end{abstract}

\maketitle

\section{Introduction}

Self-bound quantum droplets are ultradilute liquid-like states stabilized by beyond-mean-field (BMF) quantum fluctuations. They provide a versatile platform for studying nonlinear collective phenomena in ultracold atomic gases \cite{Tarruellscience,Pfau2016,Fattori2018,Chomaz_2023}. In a binary Bose mixture, the mean-field (MF) theory predicts collapse when the attractive interspecies interaction overcomes the repulsive intraspecies interaction \cite{Petrov,Malomed2018}. The collapse can be arrested by the Lee-Huang-Yang (LHY) correction \cite{LHY1957}, whose repulsive BMF contribution balances the attractive MF energy and gives rise to a self-bound state. Depending on the interaction parameters and atom number, one may distinguish solitonic, crossover, and bulk-like droplet regimes. 
The same MF-BMF competition also supports different localized structures, including bright droplets \cite{Tarruellscience}, dark droplets \cite{Matthew22}, and soliton-droplet hybrids \cite{Kevrekidis23}.

Droplet physics has developed rapidly in recent years. Experiments and theory have explored collective modes \cite{Dmitry2020}, quenches and nonequilibrium dynamics \cite{Sadeghpour2022,Bisset2022,Yang2023}, collisions \cite{Fattori2019,Boronat2021,Hu2022}, and the dynamical formation of multiple droplets in Bose-Bose mixtures \cite{Burchianti2025}. Recent theoretical work has further extended the concept to curved geometries, where LHY stabilization can support droplet states on spherical surfaces \cite{Li2025SphericalDroplets}, and to asymmetric one-dimensional Bose-Bose mixtures, where unequal intraspecies couplings strongly modify the density profile and excitation spectrum \cite{Xiao2026Asymmetric1D}. These developments highlight the importance of having reliable tools for steering droplet configurations in a controlled and time-efficient way.

A central challenge is to achieve fast state-to-state manipulation of self-bound droplets. Ideally, one would like to connect two stationary droplet configurations within a short time while avoiding breathing excitations, shape deformations, or loss of stability. This is nontrivial because the one-dimensional droplets considered
here rely on a delicate balance between residual repulsive MF
interactions and an attractive BMF contribution; rapid interaction
changes can drive the system away from its instantaneous equilibrium.

Shortcuts to adiabaticity (STA) \cite{ReviewSTA} offer a route to overcome this limitation. STA methods are designed to reproduce the final outcome of an adiabatic transformation without requiring slow evolution. In ultracold gases, related inverse-engineering ideas have been used to control nonlinear matter waves \cite{Li2016} and few- and many-body systems \cite{Ruks:19,Kahan:19,Hasan2024}. These results suggest that interaction-based STA protocols can also be useful for the controlled manipulation of self-bound quantum droplets.

A suitable variational description is essential for constructing such protocols. Gaussian ansatzes have been widely used to model droplet dynamics \cite{Abdullaev20,Fatkhulla2019,Reimann2021,Mistakidis2025,Fatkhulla2026}, but they do not accurately capture the flat-top profiles that appear in high-norm droplets. Analytical bright- and dark-droplet profiles \cite{Matthew22} provide a more appropriate starting point across different droplet regimes. Motivated by this, we use a variational ansatz capable of describing both small-norm and large-norm configurations \cite{Malomed2018,Kevrekidis23}, and combine it with inverse engineering to obtain explicit time-dependent interaction protocols.

In this work, we develop a variational inverse-engineering method for
the rapid control of quantum droplets within an effective
one-dimensional extended Gross--Pitaevskii description, using
time-dependent MF and BMF nonlinearities as control parameters.
Our goal is to connect prescribed initial and target droplet states with high final fidelity and minimal residual excitation. We first consider the unconstrained case, where both nonlinearities {can be controlled simultaneously in time}. We then examine two experimentally motivated constrained settings:
one with a fixed BMF nonlinearity and a time-dependent MF interaction,
and the other with a fixed MF nonlinearity and a time-dependent BMF
interaction.
In addition, we quantify the sensitivity of the resulting protocols to finite calibration errors in the nonlinear control strength.

The manuscript is organized as follows. In Sec.~\ref{sec:2}, we present the effective one-dimensional model for a binary Bose-Einstein condensate with tunable interactions. In Sec.~\ref{sec:3}, we introduce the variational framework and the inverse-engineering strategy used to construct the STA protocols. In Sec.~\ref{sec:4}, we present three representative examples, discuss
their control performance, analyze the sensitivity to finite parameter variations, and briefly comment on the experimental relevance of the parameter regime. Finally, in Sec.~\ref{sec:6}, we summarize the results and discuss possible extensions.

\section{Model}
\label{sec:2}

We consider a binary BEC in one spatial dimension with mutually
symmetric spinor components under the influence of BMF corrections
that account for quantum fluctuations \cite{LowD2016,Pathak2022}.
We assume that the intraspecies couplings describing the repulsion
between atoms in the two components are equal,
$g_s\equiv\sqrt{g_{\uparrow\uparrow}g_{\downarrow\downarrow}}
=g_{\uparrow\uparrow}=g_{\downarrow\downarrow}$, and we denote the
interspecies coupling by $g_c=g_{\uparrow\downarrow}$.
The intra- and interspecies scattering lengths,
$a=a_{\uparrow\uparrow}=a_{\downarrow\downarrow}$ and
$a_{\uparrow\downarrow}$, can be experimentally tuned by Feshbach
resonances \cite{Feshbach}.

In the symmetric setting considered here, the two component fields
satisfy $\phi_1=\phi_2=\psi/\sqrt{2}$, so that $|\psi|^2$ denotes
the total linear density.
The time-dependent extended Gross--Pitaevskii equation for the
effective wave function $\psi$ in 1D is then
\cite{LowD2016,Pathak2022,Abdullaev20}
\begin{equation}
i\hbar \frac{\partial \psi}{\partial t}
=
-\frac{\hbar^2}{2m}\frac{\partial^2\psi}{\partial x^2}
+\frac{1}{2}\delta g(t)|\psi|^2\psi
-\frac{\sqrt{m}}{\pi\hbar}g(t)^{3/2}|\psi|\psi.
\label{eq:cGPE}
\end{equation}
Here $\delta g(t)=g_s(t)+g_c(t)$ and $g(t)=g_s(t)$
parametrize the MF and BMF nonlinearities, respectively, and
$m$ is the atomic mass of each component.
The dimension-specific BMF term is attractive and is balanced
by the residual repulsive MF interaction.
The effective wave function is normalized as
$\int_{-\infty}^{\infty} dx\,|\psi|^2=N_p$,
where $N_p$ is the total physical particle number of the
two-component mixture. Each component therefore contains
$N_p/2$ particles.

Let us introduce a fixed length unit $\ell$ and rescale the
coordinate as $x=\ell\tilde{x}$.
The corresponding energy and time units are
$\epsilon\equiv\hbar^2/(m\ell^2)$ and
$\tau\equiv m\ell^2/\hbar$, respectively, with $t=\tau\tilde{t}$.
We also rescale the wave function as
$\psi=\tilde{\psi}\sqrt{N_p/(\ell N)}$, where $N$ is the chosen
dimensionless normalization of the rescaled field.
In addition, we write
$\delta g=\epsilon\ell r\,\widetilde{\delta g}$ and
$g=\epsilon\ell r^{1/3}\widetilde g$, with $r=N/N_p$.
Then, Eq.~\eqref{eq:cGPE} can be rewritten in the following
dimensionless form:
\begin{equation}
i\frac{\partial\tilde{\psi}}{\partial\tilde{t}}
=
-\frac{1}{2}\frac{\partial^2\tilde{\psi}}{\partial\tilde{x}^2}
+\frac{1}{2}\widetilde{\delta g}(\tilde{t})
|\tilde{\psi}|^2\tilde{\psi}
-\frac{1}{\pi}\widetilde g(\tilde{t})^{3/2}
|\tilde{\psi}|\tilde{\psi}.
\label{eq:GPE}
\end{equation}
The dimensionless normalization is
$\int_{-\infty}^{\infty} d\tilde{x}\,|\tilde{\psi}|^2=N$.

For positive $g$ and $\delta g$, we define the dimensionless
shape parameter
\[
q=
\frac{\pi}{\sqrt{2}}N_p
\left(\frac{\delta g}{g}\right)^{3/2}
=
\frac{\pi}{\sqrt{2}}N
\left(\frac{\widetilde{\delta g}}{\widetilde g}\right)^{3/2},
\]
which is invariant under the above rescaling.
For stationary droplets, small values of $q$ correspond to
soliton-like profiles without a well-developed plateau, whereas
large values correspond to liquid-like profiles with a pronounced
flat top \cite{Malomed2018}.
The parameter $q$ characterizes the stationary profile shape
rather than its absolute spatial width.
In the following, we use Eq.~\eqref{eq:GPE} as the starting point
and omit the tildes for notational simplicity.

\section{Variational inverse-engineering shortcut scheme}
\label{sec:3}

We now show how to enable rapid and precise control of self-bound
quantum droplets by dynamically tuning their nonlinear interactions
via Feshbach resonances.
We develop optimised STA protocols using variational inverse engineering to steer transitions between droplet states while suppressing residual excitations.

\subsection{General case}

We aim to control the nonlinear coupling strengths 
$\delta g(t)$ and $g(t)$ to achieve rapid manipulation from the ground state to the desired target state while avoiding  excitations of the final state.
The initial and target values for these time-dependent strengths are set as:
\begin{eqnarray}
g(0)&=&g_{0}, \quad g(t_f)=g_{f}, \nonumber \\
 \delta g(0)&=&\delta g_{0}, \quad  \delta g(t_f)=\delta g_{f},
\label{eq:g1g2condi}
\end{eqnarray}
where $g_0$ and $\delta g_0$
denote the initial values of the coupling strengths, while 
$g_f$ and $\delta g_f$ represent their target values at the final time $t_f$.
The dynamical variational approach is employed \cite{Zoller1997} using the following trial function:
\begin{equation}
\label{eq:trial}
\psi(x,t) = \sqrt{Ns(t)}\,u\left(y(t), \eta(t) \right) \exp\left[\mathrm{i} b(t) x^2 + \mathrm{i} \varphi(t) \right],
\end{equation}
where $b(t)$, $s(t)$, $\eta(t)$, and $\varphi(t)$ are real
time-dependent variational parameters describing the chirp, inverse
width, shape parameter, and global phase, respectively. 
The function $u(y(t),\eta(t))$ represents the real part of the wave packet, where $y=x s(t)$. For the sake of simplicity, we omit the variable $t$ of these functions.
With the dimensionless normalization
$\int dx\,|\psi(x,t)|^2=N$, the ansatz in Eq.~(\ref{eq:trial})
requires
\begin{equation}
\int_{-\infty}^{+\infty} dy\,|u(y,\eta)|^2=1 .
\label{eq:norm}
\end{equation} 
We use the following ansatz for $u(y,\eta)$:
\begin{equation}
    u(y,\eta) = \frac{a}{1+\eta \cosh{y}}.
\end{equation}
The normalization {condition} for $u(y)$ in Eq. (\ref{eq:norm}) is 
\begin{equation}
    \label{eq:unorm}
    \frac{2 a^2 \left(\sqrt{\eta ^2-1}-2 \chi\right)}{\left(\eta ^2-1\right)^{3/2}}=1,
\end{equation}
where $\chi = \tan^{-1}\left(\frac{\eta +1}{\sqrt{\eta ^2-1}}\right)- \cot ^{-1}\left(\sqrt{\eta ^2-1}\right)$.
In the limit $\eta\rightarrow 0$, the ansatz approaches a flat-top
density profile with increasingly sharp edges.
The function $a$ can be determined by the normalization condition Eq.~(\ref{eq:unorm}).

We substitute the ansatz Eq.~\eqref{eq:trial} into the Lagrangian $L=\int_{-\infty}^{+\infty} dx \mathcal{L} $ for the effective 1D model of Eq.~\eqref{eq:GPE}, where
\begin{align}
    \label{eq:fulllag}
    &&\mathcal{L}\left[ \psi, \psi^*\right] 
    = \frac{i}{2} \left(\psi \psi^*_{t} - \psi^* \dot{\psi} \right)+E(\psi),\\
    &&E(\psi)=\frac{1}{2}\left\vert \frac{\partial \psi}{\partial x} \right\vert^2 -\frac{2}{3\pi}g^{3/2} \vert \psi \vert^3+ \frac{1}{4} \delta g\vert \psi \vert^4.
    \label{eq:energy}   
\end{align}
Here a star denotes the complex conjugate and the subscript $t$ denotes the derivative with respect to time. 
After substituting the ansatz and dividing the integrated Lagrangian by
$N$, we obtain the effective Lagrangian
\begin{eqnarray}
L =\frac{\dot{b} + 2b^2}{s^2} w^2+ \dot{\varphi}+ s^2 E_{\rm kin}^u+\frac{\sqrt{sN}}{\pi}g^{3/2} E_{\rm BMF}^u +\frac{sN}{2}\delta g  \,E_{\rm MF}^u .\nonumber \\
\label{eq:fullL}
\end{eqnarray}
The integrals $w(\eta)$, $E_{\rm kin}^u(\eta)$, $E_{\rm BMF}^u(\eta)$ and $E_{\rm MF}^u(\eta)$ correspond to the width, kinetic, BMF and MF energies, which depend only on the variable $\eta$. 
They are defined by
\begin{eqnarray}
\label{eq:wfgk}
&& w (\eta)=\sqrt{\int dy\; y^2 \vert u \vert^2}, \quad\quad \;E_{\rm kin}^u=\frac{1}{2}\int dy\;\big\vert \frac{\partial u}{\partial y}\big\vert^2,\; \nonumber\\
&& E_{\rm BMF}^u(\eta)=-\frac{2}{3}\int dy\; \vert u \vert^3, \quad  E_{\rm MF}^u(\eta)=\frac{1}{2}\int dy\; \vert u \vert^4,
\label{eq:energy_u}
\end{eqnarray}
where we have omitted the parameter $\eta$ inside the integrals for the sake of simplicity.

The dynamics of quantum droplets for some generalised coordinates $q_i=\{\eta,s,b,\varphi\}$, emerge from the respective Euler-Lagrange equations $\partial L/\partial q_i = d\left( \partial L/\partial \dot{q_i}\right)/dt$.
Applying this for the variable $b$, one can express $b$ in terms of the width $\sigma=w/s$ of the quantum droplet 
and obtain
\begin{eqnarray}
\label{eq:b}
    b(t)=\frac{\dot{w}}{2w}-\frac{\dot{s}}{2s}\equiv\frac{\dot{\sigma}}{2\sigma}.
\end{eqnarray}
The chirp $b$ is well known to be related to the width of the wavepacket \cite{Abdullaev20,otajonov2024}. 
After applying $\partial L/\partial s =0$ and combining with Eq.~\eqref{eq:b}, we get an Ermakov-like equation \cite{Li2016}
\begin{equation}
    \ddot{\sigma}=\frac{1}{\sigma}\left[2s^2 E_{\rm kin}^u+\frac{1}{2\pi}\sqrt{Ns}g^{3/2} E_{\rm BMF}^u+\frac{Ns}{2}\delta g \, E_{\rm MF}^u\right].\label{eq:dLds}
\end{equation}
It shows that the width of the droplet is associated with the strengths of the BMF and MF interactions. Applying
$\partial L/\partial\eta=0$ gives
\begin{align}
&\frac{2w(\dot{b}+2b^2)}{s^2}\partial_\eta w
+s^2\partial_\eta E_{\rm kin}^u
\nonumber \\
&\quad
+\frac{\sqrt{sN}}{\pi}g(t)^{3/2}
\partial_\eta E_{\rm BMF}^u
+\frac{sN}{2}\delta g(t)\partial_\eta E_{\rm MF}^u
=0 .
\label{eq:dLdeta}
\end{align}
Here $\partial_\eta$ denotes differentiation with respect to $\eta$,
and the energy terms are those defined in Eq.~(\ref{eq:energy_u}).

Let us now consider a stationary adiabatic state with the width being time independent, so that
$\dot{\sigma}=0$ and $\ddot{\sigma}=0$. Substituting
$\ddot{\sigma}=0$ into Eq.~\eqref{eq:dLds} and introducing
$m_0=\sqrt{s_0}$ gives
\begin{equation}
2E_{\rm kin}^u m_0^3
+
\frac{N}{2}\delta g E_{\rm MF}^u m_0
+
\frac{\sqrt{N}}{2\pi}g^{3/2}E_{\rm BMF}^u
=0 .
\label{eq:m0}
\end{equation}
For a given value of the shape parameter $\eta$, Eq.~\eqref{eq:m0}
determines the stationary scale parameter through $s_0=m_0^2$, where
the physically relevant real root is selected in the numerical
implementation. The stationary value of $\eta$ is obtained from the
Euler--Lagrange equation for the shape parameter. In a stationary state,
$b=\dot b=0$, and Eq.~\eqref{eq:dLdeta} reduces to an algebraic
condition for $\eta_0$ at fixed $g$ and $\delta g$. Equations
\eqref{eq:m0} and \eqref{eq:dLdeta} are therefore solved together to
obtain the stationary values of $s$ and $\eta$ used in the initial and
target variational wave functions.

According to the above analysis, the key ingredients are the integrals $w(\eta)$, $E_{\rm kin}^u(\eta)$, $E_{\rm BMF}^u(\eta)$ and $E_{\rm MF}^u(\eta)$ in Eq.~\eqref{eq:wfgk}.

\begin{figure}[t]
\centering
\includegraphics[width=0.86\columnwidth]{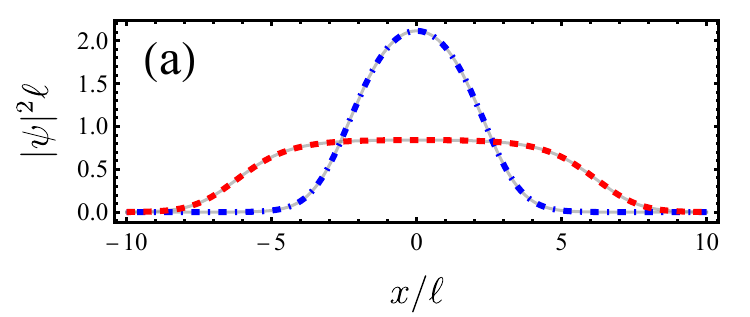}
\includegraphics[width=0.86\columnwidth]{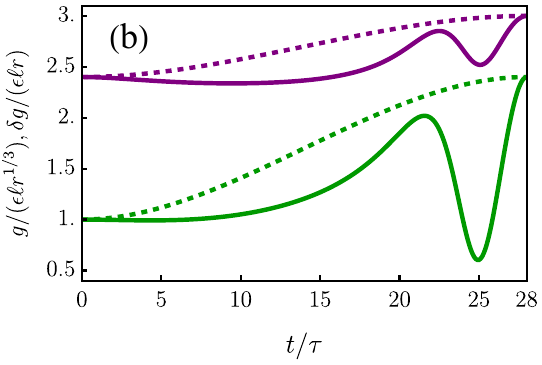}
\includegraphics[width=0.86\columnwidth]{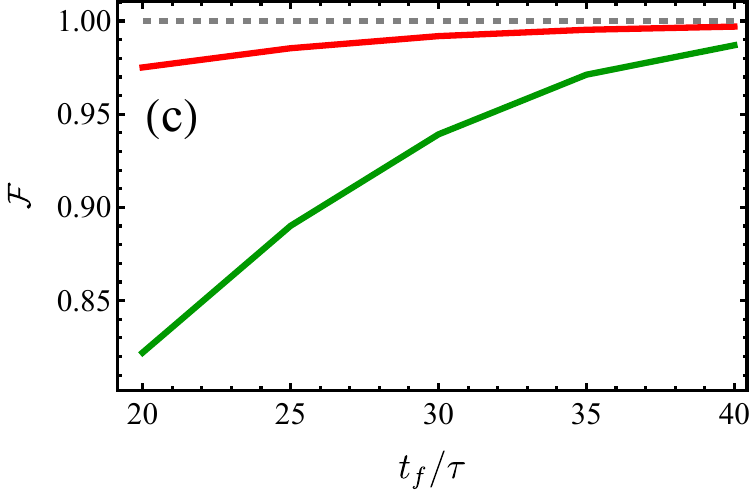}
\caption{Case I: STA with both nonlinearities varied in time.
(a) Initial density profile (blue dash-dotted curve) and target density
profile (red dashed curve), as given by Eq.
(\ref{eq:target}). The corresponding numerical ground-state densities
(light gray) agree well with the variational ansatz.
{(b) Time-dependent nonlinearities for the BMF coefficient
$g(t)/(r^{1/3}\epsilon\ell)$ (purple) and the MF coefficient
$\delta g(t)/(r\epsilon\ell)$ (green). The solid curves are the
STA-designed protocols obtained from Eqs.~(\ref{eq:g-control}) and
(\ref{eq:deltag-control}), while the dotted curves are the reference
ramps defined in Eq.~(\ref{eq:refgt}). The boundary values are
$g_0/(r^{1/3}\epsilon\ell)=2.4$,
$g_f/(r^{1/3}\epsilon\ell)=3$,
$\delta g_0/(r\epsilon\ell)=1$, and
$\delta g_f/(r\epsilon\ell)=2.4$.}
(c) Final fidelity $\mathcal{F}$ as a function of the final time
$t_f/\tau$ for the STA protocol (red) and the reference ramp (green).
The gray dotted line indicates the ideal value $\mathcal{F}=1$. The
STA protocol reaches the high-fidelity regime, $\mathcal{F}>0.99$, for
$t_f/\tau \gtrsim 28$. The norm is $N=10$.
}
\label{fig:case0}
\end{figure}

The initial and target states can be written as
\begin{equation}
\psi_0(x)=
\frac{A_0}{1+\eta_0\cosh(s_0x)},
\qquad
\psi_f(x)=
\frac{A_f}{1+\eta_f\cosh(s_fx)}, \label{eq:target}
\end{equation}
where $A_{0,f}=\sqrt{Ns_{0,f}}\,a_{0,f}$.
For example, the initial and final target states are illustrated in Fig.~\ref{fig:case0}(a). 
Our goal is to efficiently control the system's dynamics to ensure a smooth transition between these two states.

The strategy for designing the STA protocols for $g(t)$ and $\delta g(t)$ is derived inversely from Eq.~\eqref{eq:dLds} and Eq.~\eqref{eq:dLdeta} by specifying the functions $s(t)$ and $\eta(t)$. 
To ensure smooth dynamics, we impose six additional boundary conditions on both $s(t)$ and $\eta(t)$:
\begin{align}
\label{eq:bounds-1}
&\dot{s}(0)=\dot{s}(t_f)=\ddot{s}(0)=\ddot{s}(t_f)=\dddot{s}(0)=\dddot{s}(t_f)=0, \\
&\dot{\eta}(0)=\dot{\eta}(t_f)=\ddot{\eta}(0)=\ddot{\eta}(t_f)=\dddot{\eta}(0)=\dddot{\eta}(t_f)=0.
\label{eq:bounds-2}
\end{align}
These 
guarantee that the initial and final states are adiabatic correspondences for designing STA protocols.
Having fixed the boundary conditions, the trajectories of $s(t)$ and $\eta(t)$
can be chosen, e.g. using  
the polynomial ansatz of the form 
\begin{equation}
  s(t)=\sum_{n=0}^{l} s_{n} t^n, \quad \eta(t)=\sum_{n=0}^{m} \eta _n t^n,
  \label{eq:polys}
\end{equation}
where the number of boundary conditions determines $l$ and $m$.

{Substituting the auxiliary trajectories $s(t)$ and $\eta(t)$ into
Eqs.~(\ref{eq:dLds}) and (\ref{eq:dLdeta}), the two nonlinear control
functions $g(t)$ and $\delta g(t)$ can be obtained algebraically. For
compactness, we define
\begin{align}
    R_s(t)
    &=
   \sigma \ddot{\sigma}
    -
    2s^2 E_{\rm kin}^u ,
    \\
    R_\eta(t)
    &=
    -
    \left[
    \frac{2w(\dot{b}+2b^2)}{s^2}\partial_\eta w
    +
    s^2\partial_\eta E_{\rm kin}^u
    \right],
\end{align}
where $\sigma=w/s$ and $b=\dot{\sigma}/(2\sigma)$. All
$\eta$-dependent quantities are evaluated at $\eta=\eta(t)$.
Solving Eqs.~(\ref{eq:dLds}) and (\ref{eq:dLdeta}) for
$g(t)^{3/2}$ and $\delta g(t)$ then gives
\begin{align}
    g(t)^{3/2}
    &=
    \frac{sN}{2\mathcal{D}}
    \left[
    \left(\partial_\eta E_{\rm MF}^u\right)R_s
    -
    E_{\rm MF}^u R_\eta
    \right],
    \label{eq:g-control}
    \\
    \delta g(t)
    &=
    \frac{\sqrt{sN}}{\pi\mathcal{D}}
    \left[
    \frac{1}{2}E_{\rm BMF}^u R_\eta
    -
    \left(\partial_\eta E_{\rm BMF}^u\right)R_s
    \right],
    \label{eq:deltag-control}
\end{align}
with
\begin{equation}
    \mathcal{D}
    =
    \frac{sN\sqrt{Ns}}{4\pi}
    \left[
    E_{\rm BMF}^u\partial_\eta E_{\rm MF}^u
    -
    2E_{\rm MF}^u\partial_\eta E_{\rm BMF}^u
    \right].
    \label{eq:control-denominator}
\end{equation}
Equations~(\ref{eq:g-control}) and (\ref{eq:deltag-control}) show
explicitly that the STA protocols are fully determined once the
auxiliary functions $s(t)$ and $\eta(t)$ are specified.}

\subsection{Incorporating constraints\label{constraints}}

We now consider the case if there are additional constraints on the controllability of the
nonlinearities, for example if the BMF nonlinearity needs to be
kept constant, $g(t)=g$, while only the MF nonlinearity $\delta g(t)$
can be varied in time or the other way around.


A key aspect of our approach is the significant flexibility available in selecting the auxiliary functions $s(t)$ and $\eta(t)$. 
Our strategy is to {use this freedom to satisfy the imposed constraints as closely as possible, without changing the STA boundary conditions at $t=0$ and $t=t_f$. 
Specifically, we introduce additional free parameters at interior interpolation points of the auxiliary trajectories,}
\begin{equation}
\begin{aligned}
s\left(\frac{i t_f}{L}\right) &= s_i+c_i,
\qquad i=1,\ldots,L-1,\\
\eta\left(\frac{j t_f}{M}\right) &= e_j+r_j,
\qquad j=1,\ldots,M-1 .
\end{aligned}
\label{eq:cs}
\end{equation}
Here $s_i$ and $e_j$ are the corresponding values of {the baseline polynomial trajectories in Eq. \eqref{eq:polys} evaluated at the interior times $i t_f/L$ and $j t_f/M$, respectively. The integers $L$ and $M$ specify the number of subintervals used to place these interior points, and therefore the numbers of additional free parameters are $L-1$ for $s(t)$ and $M-1$ for $\eta(t)$. 
We then
optimize the parameters $c_i$ and $r_j$ so that the resulting control
functions satisfy the imposed constraint as closely as possible.}


In the following, we will present three examples to detail our strategies for the protocol optimisations.

\begin{figure}[t]
\includegraphics[width=0.86\columnwidth]{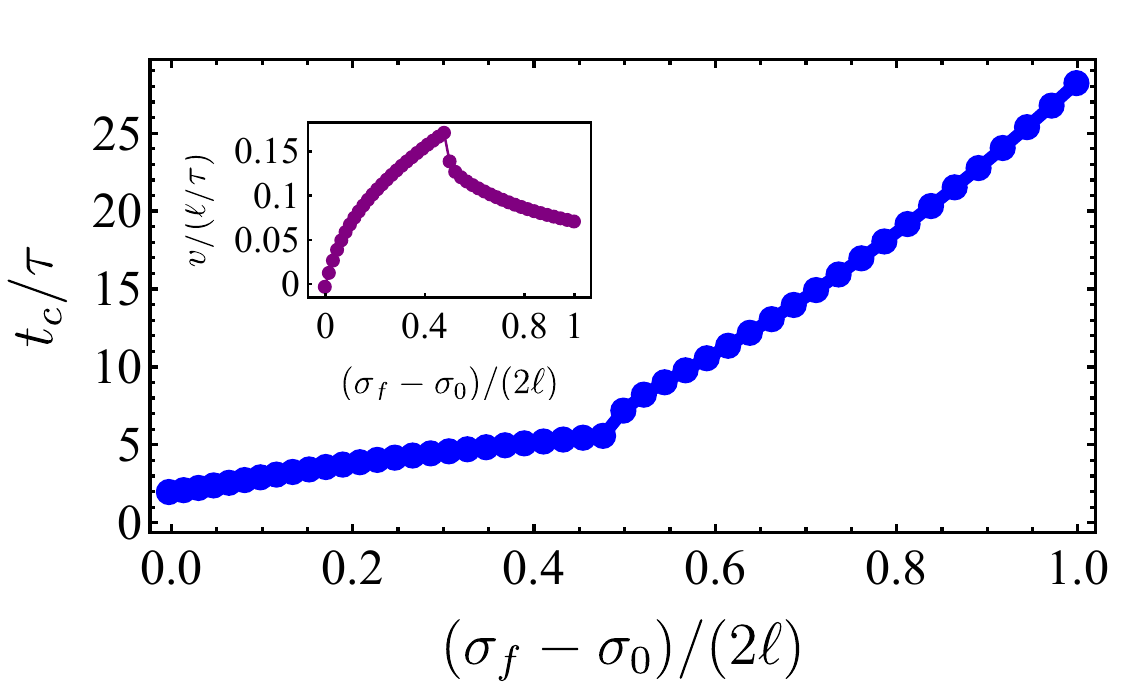}
\caption{{Critical time for high-fidelity expansion in Case I. The critical final
time $t_c/\tau$, defined by the condition $\mathcal{F}>0.99$, is shown
as a function of the normalized width difference
$(\sigma_f-\sigma_0)/(2\ell)$ between the target and initial droplets.
The parameters are $g_0/(r^{1/3}\epsilon\ell)=2.4$,
$g_f/(r^{1/3}\epsilon\ell)=3$, and
$\delta g_0/(r\epsilon\ell)=1$, while
$\delta g_f/(r\epsilon\ell)$ is varied from $1.48$ to $2.4$. For
$N=10$, the corresponding flatness parameter changes from
$q_0\simeq 5.97$ initially to $q_f\simeq7.7$--$15.90$ for the target
states. The inset shows the average expansion velocity
$v/(\ell/\tau)=(\sigma_f-\sigma_0)/t_c$.
}
}
\label{fig:case0_vel}
\end{figure}

\section{Example Cases}
\label{sec:4}

To evaluate the effectiveness of the STA protocols, we calculate
the fidelity at the final time $t_f$,
\begin{equation}
\mathcal{F}
=
\frac{
\left|\int dx\,\Psi_f^*(x)\psi(x,t_f)\right|^2
}{
\left(\int dx\,|\Psi_f(x)|^2\right)
\left(\int dx\,|\psi(x,t_f)|^2\right)
},
\label{eq:fidelity}
\end{equation}
where $\Psi_f$ is the numerical ground state at the target
interaction strengths, obtained by imaginary-time evolution  \cite{BaoDu2004},
and $\psi(x,t_f)$ is the state reached under the applied protocol.
As a reference for comparison, we use smooth polynomial ramps for
$g(t)$ and $\delta g(t)$ that satisfy the initial and final boundary
conditions, 
\begin{equation}
    g^{(R)}_{j}(t) = g_{j,0} - 3\left(g_{j,0} - g_{j,f}\right)\left(\frac{t}{t_f}\right)^2 + 2\left(g_{j,0} - g_{j,f}\right)\left(\frac{t}{t_f}\right)^3.
    \label{eq:refgt}
\end{equation} 
Here {$g^{(R)}_j(t)$ denotes the protocol of $g^{(R)}_1(t)\equiv g(t)$ and $g^{(R)}_2(t)\equiv \delta g(t)$} for the reference ramps. The initial $g_{j,0}$ and final values $g_{j,f}$ are given by Eq.~\eqref{eq:g1g2condi}.

In the following, we will study three examples.
{In Case I, we will design time-dependent $g(t)$ and $\delta g(t)$ control functions
without any constraints}. Then, two extreme constraints are examined:
first, {as Case II,} the constraint of a constant $g$ while $\delta g(t)$ only depends on time and then, {as Case III}, the constraint of a constant $\delta g$ where $g(t)$ depends only on time.

\subsection{Case I: Time-dependent $\delta g (t)$ and $g(t)$} 

We assess the validity of the ansatz presented in Eq.~\eqref{eq:trial} by conducting a comparison with the exact ground state.  
To achieve this, we utilize the technique of imaginary time evolution \cite{BaoDu2004} to calculate the ground state wave function, which we denote as the initial state $\Psi_{0}$ and the ideal target state $\Psi_f$. 
{In the following numerical examples, the nonlinear coefficients are
reported in the dimensionless rescaled units introduced in
Sec.~\ref{sec:2}. Specifically, the plotted and quoted values correspond
to
$\tilde g=g/(\epsilon\ell r^{1/3})$
and
$\widetilde{\delta g}=\delta g/(\epsilon\ell r)$.
When discussing the control protocols, we keep the notation $g$ and
$\delta g$ to emphasize their connection to the MF and BMF interaction
strengths controlled through the scattering lengths.}

{As a representative example, we consider an expansion from a relatively
narrow localized droplet to a broader droplet with a more pronounced
flat-top density profile.
The initial parameters are
$g_0/(r^{1/3}\epsilon\ell)=2.4$ and
$\delta g_0/(r\epsilon\ell)=1$ for $N=10$. The target state is obtained with
$g_f/(r^{1/3}\epsilon\ell)=3$ and
$\delta g_f/(r\epsilon\ell)=2.4$, corresponding to
$q_f\simeq15.90$. Thus, the protocol connects an initially localized
droplet to a final state with a more pronounced flat-top density
profile.}
In Fig. \ref{fig:case0}(a), we illustrate the comparison between the variational ansatz $\psi_0$ (blue dash-dotted line) and $\psi_f$ (red dashed line) with fixed stationary parameters $(\eta_{0,f}, s_{0,f}, b_{0,f}, \varphi_{0,f})$ 
and the numerical ground state (gray lines) $\Psi_{0,f}$, showing excellent agreement. 

Furthermore, we present the STA {control functions} (solid lines) $g(t)/(r^{1/3}\epsilon\ell)$ (purple) and $\delta g(t)/(r\epsilon\ell)$ (green) {for final time $t_f/\tau = 28$, in Fig. \ref{fig:case0} (b).
The corresponding fidelities achieved by the STA control schemes for different final times $t_f$ are shown in
Fig. \ref{fig:case0}(c) (red, solid line). 
High fidelities are achieved, specifically, a fidelity $\mathcal{F}>0.99$ is obtained for final times $t_f$ longer than a critical final time $t_c/\tau \approx 28$. We also show that the performance of the designed STA is significantly higher than the fidelity of the reference case (green, solid line in Fig. \ref{fig:case0}(c)).}

To characterize the density profile of the droplet, we recall the
dimensionless flatness parameter $q = \pi/\sqrt{2} \left(\delta g/g\right)^{3/2} N_p$,
where larger values of $q$ correspond to a more pronounced flat-top
profile. In Fig.~\ref{fig:case0_vel}, we keep
$g_0/(r^{1/3}\epsilon\ell)=2.4$, and
$\delta g_0/(r\epsilon\ell)=1$ fixed, resulting in an initial state with a width $\sigma_0$.
We also keep $g_f/(r^{1/3}\epsilon\ell)=3$ fixed, while varying
$\delta g_f/(r\epsilon\ell)$ from $1.48$ to $2.4$. For $N=10$, this
corresponds to a fixed initial value $q_0\simeq5.97$ and final values
$q_f\simeq7.7$--$15.90$. Thus the variation of $\delta g_f$ generates
a family of target droplets with different width $\sigma_f$.
Figure~\ref{fig:case0_vel} shows the critical time $t_c/\tau$ required
to reach $\mathcal{F}>0.99$ as a function of the normalized width
difference $(\sigma_f-\sigma_0)/(2\ell)$. The critical time increases
rapidly as the target droplet becomes wider and more flat-top-like.
The inset shows the corresponding average expansion velocity,
$v/(\ell/\tau)=(\sigma_f-\sigma_0)/t_c$. Its maximum reflects the
competition between the increasing width change and the faster growth
of the critical time required to reach broader, more flat-top target
droplets while maintaining high final fidelity.

\subsection{Case II: Constant $g$ and time-dependent $\delta g(t)$}


\begin{figure}[t]
\includegraphics[width=0.75\columnwidth]{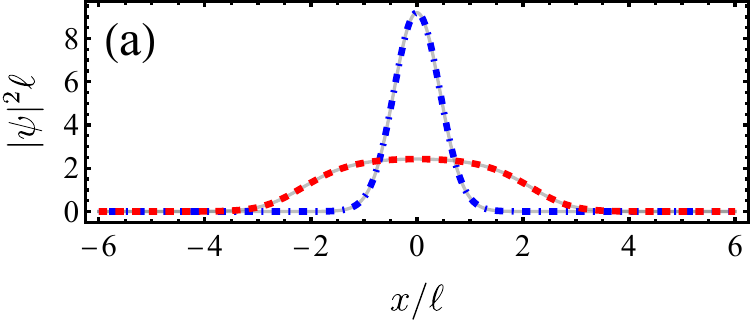}
\includegraphics[width=0.79\columnwidth]{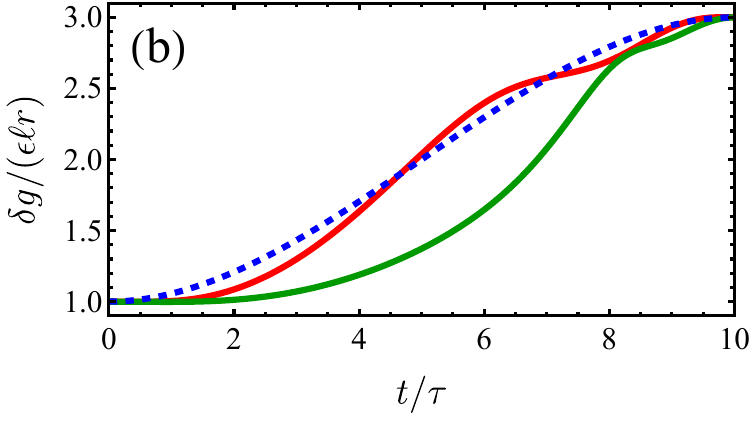}
\includegraphics[width=0.91\columnwidth]{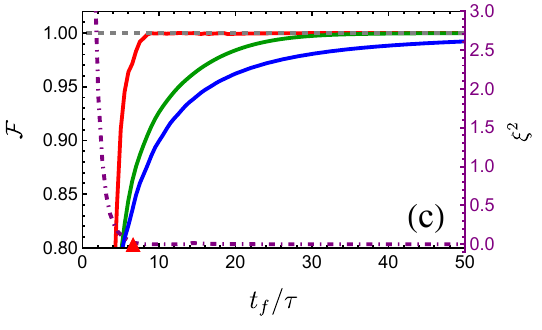}
\hspace{-9mm}
\caption{
Case II: constrained expansion with fixed BMF nonlinearity.
(a) Initial and target density profiles.
(b) MF nonlinearity $\delta g(t)/(\epsilon\ell r)$ for the reference
ramp (dotted blue), forced STA protocol (green), and optimised STA
protocol (red), shown for $t_f/\tau=10$.
Here $\delta g_0/(\epsilon\ell r)=1$,
$\delta g_f/(\epsilon\ell r)=3$, and
$g/(\epsilon\ell r^{1/3})=5$ is kept constant.
(c) Fidelity $\mathcal{F}$ (left axis) and constraint error $\xi^2$
(purple dotted curve, right axis) as functions of $t_f/\tau$.
{The triangle marks the shortest final time at which the optimised STA
protocol simultaneously satisfies $\mathcal{F}>0.99$ and
$\xi^2\leq 5\times10^{-2}$. The norm is $N=10$.}
}
\label{fig:caseA}
\end{figure}

We next consider the constrained expansion protocol in which the BMF
nonlinearity is kept constant, $g(t)=g_{\rm c}$, while only the MF
nonlinearity $\delta g(t)$ is used as the time-dependent control.
{The corresponding initial and final density profiles are shown in Fig. \ref{fig:caseA}(a), showing again a very good agreement between the ansatz and the exact ground states.}

{The first control strategy is the forced STA protocol. We first
calculate both time-dependent nonlinearities as in the unconstrained
Case I, and then impose the Case-II constraint by hand by setting
$g(t)=g_{\rm c}=g_0=g_f$. The corresponding STA-designed
$\delta g(t)$ is kept as the control function. This protocol is shown
as the green curve in Fig.~\ref{fig:caseA}(b) for $t_f/\tau=10$.
The resulting fidelity, shown in Fig.~\ref{fig:caseA}(c), improves
substantially compared with the reference ramp, but reaches
$\mathcal{F}>0.99$ only for relatively long final times,
$t_f/\tau\simeq 30$.

The second strategy is the optimised STA protocol. In this case, the
additional freedom in the auxiliary trajectories $s(t)$ and $\eta(t)$
is used to make the inverse-engineered BMF nonlinearity as close as
possible to the required constant value $g_{\rm c}$. For each set of
parameters $c_i$ and $r_j$, Eq.~(\ref{eq:cs}) defines modified
auxiliary trajectories, which are then substituted into the
inverse-engineering equations to obtain $g(c_i,r_j,t)$ and
$\delta g(c_i,r_j,t)$. 
The optimisation is then performed by minimising the constraint error}
\begin{equation}
    \xi^2
    =
    \frac{1}{t_f}
    \int_0^{t_f}
    \left[
    g(c_i,r_j,t)-g_{\rm c}
    \right]^2 dt .
    \label{eq:xi_caseII}
\end{equation}
For the results shown here, we use Eq.~(\ref{eq:cs}) with $L=3$ and
$M=8$, giving nine free parameters in total: two associated with
$s(t)$ and seven with $\eta(t)$.

After optimizing over $c_i$ and $r_j$, the resulting MF control
function $\delta g(c_i,r_j,t)$ is shown as the red curve in
Fig.~\ref{fig:caseA}(b). Although this optimised ramp remains close to
the reference ramp in amplitude, its small time-dependent deviations
are selected to follow an inverse-engineered trajectory that satisfies
the fixed-$g$ constraint more accurately. As shown in
Fig.~\ref{fig:caseA}(c), the optimised STA protocol strongly suppresses
residual excitations and reaches the high-fidelity regime
$\mathcal{F}>0.99$ already at $t_f/\tau\simeq 7$, while keeping the
constraint error below $\xi^2\leq 5\times10^{-2}$.

We note that the fidelity is not included directly in the cost function.
Within the variational inverse-engineering construction, a trajectory
that exactly satisfies the imposed boundary conditions is designed to
connect the initial and target variational states. The present
optimisation therefore focuses on enforcing the fixed-$g$ constraint.
A further optimisation directly targeting the full numerical fidelity
could be considered separately.

\subsection{Case III: Constant $\delta g$ and time-dependent $g(t)$} 

\begin{figure}[t]
\includegraphics[width=0.75\columnwidth]{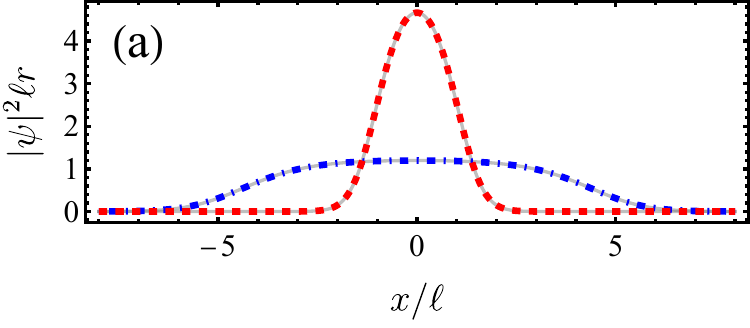}
\includegraphics[width=0.79\columnwidth]{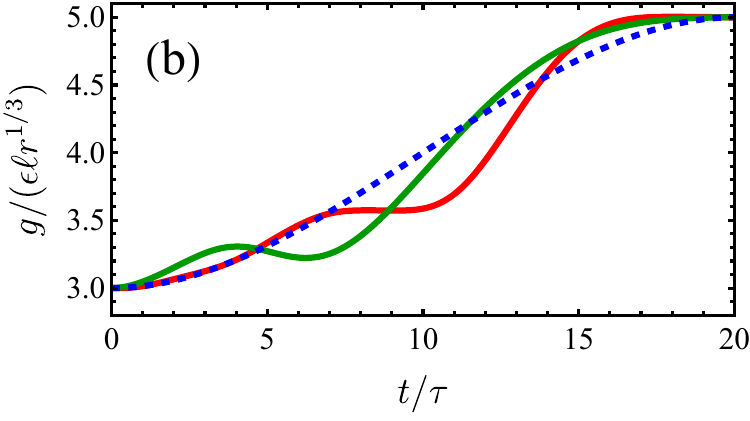}
\begin{minipage}{\linewidth}
  \centering
  \makebox[0pt][l]{\raisebox{10ex}{\hspace{60mm} \large{\large(c)}}}
\includegraphics[width=0.91\columnwidth]{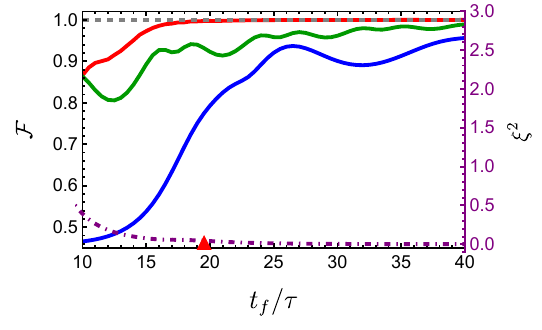}
\end{minipage}
\hspace{-9mm}
\caption{
Case III: constrained compression with fixed MF nonlinearity.
(a) Initial and target density profiles.
(b) BMF nonlinearity $g(t)/(\epsilon\ell r^{1/3})$ for the reference ramp
(dotted blue), forced STA protocol (green), and optimised STA protocol
(red). Here $g_0/(\epsilon\ell r^{1/3})=3$,
$g_f/(\epsilon\ell r^{1/3})=5$, and the MF nonlinearity is kept fixed at
$\delta g/(\epsilon\ell r)=2$.
(c) Fidelity $\mathcal{F}$ (left axis) and constraint error $\xi^2$
(purple dotted curve, right axis) versus the final time $t_f/\tau$.
The optimised STA protocol achieves $\mathcal{F}>0.99$ for
$t_f/\tau\gtrsim20$ while maintaining
$\xi^2\leq5\times10^{-2}$.
The triangle marks the shortest final time at which both criteria are
simultaneously satisfied. The norm is $N=10$.
}
\label{fig:caseB}
\end{figure}

Finally, we consider the opposite scenario, in which the MF nonlinearity is kept constant, $\delta g(t)=\delta g_{\rm c}$, while the
BMF nonlinearity $g(t)$ is used as the time-dependent control.
This case corresponds to a compression protocol. 
{The initial state is a broad
flat-top-like droplet with
$g_0/(\epsilon\ell r^{1/3})=3$ and
$\delta g_{\rm c}/(\epsilon\ell r)=2$, corresponding to
$q_0\simeq12.1$ for $N=10$. The target state has
$g_f/(\epsilon\ell r^{1/3})=5$, with the same fixed value of
$\delta g_{\rm c}$, giving $q_f\simeq5.6$. 
Since $q\propto(\delta g/g)^{3/2}$, increasing $g$ at fixed $\delta g$
reduces the flatness parameter and compresses the droplet into a
narrower, more soliton-like localized state.} The corresponding initial
and final density profiles are shown in Fig.~\ref{fig:caseB}(a), where
the variational ansatz again agrees well with the numerical ground
states.

As in Case II, we compare a forced STA protocol and an optimised STA
protocol. In the forced STA protocol, both time-dependent nonlinearities
are first obtained as in the unconstrained Case I. The Case-III
constraint is then imposed by hand by setting
$\delta g(t)=\delta g_{\rm c}=\delta g_0=\delta g_f$, and the
corresponding STA-designed $g(t)$ is kept as the control function.

For the optimised STA, the additional freedom in the auxiliary
trajectories $s(t)$ and $\eta(t)$ is used to make the inverse-engineered
MF nonlinearity as close as possible to the imposed constant value
$\delta g_{\rm c}$. For each set of parameters $c_i$ and $r_j$,
Eq.~(\ref{eq:cs}) defines modified auxiliary trajectories. Substituting
these trajectories into the inverse-engineering equations gives the
corresponding functions $g(c_i,r_j,t)$ and
$\delta g(c_i,r_j,t)$. The optimisation minimizes the constraint error
\begin{equation}
    \xi^2
    =
    \frac{1}{t_f}
    \int_0^{t_f}
    \left[
    \delta g(c_i,r_j,t)-\delta g_{\rm c}
    \right]^2 dt .
\end{equation}
Here $\delta g_{\rm c}$ is the imposed constant value. In the
calculations shown here, we use the same choice as in Case II, namely
$L=3$ and $M=8$, giving nine additional free parameters in total: two
associated with $s(t)$ and seven with $\eta(t)$. This choice provides
sufficient flexibility to reduce the constraint error below the target
threshold without introducing unnecessary additional parameters.

The comparison between the BMF nonlinearity $g(t)$ for the optimised
STA protocol, the forced STA protocol, and the reference ramp is shown
in Fig.~\ref{fig:caseB}(b). The corresponding fidelities are shown in
Fig.~\ref{fig:caseB}(c). 
The forced STA protocol improves the fidelity
compared with the reference ramp, but the pronounced oscillations in
$\mathcal{F}$ as a function of $t_f$ indicate that residual collective
excitations are still generated when the constraint is imposed by hand.
{The optimised STA protocol suppresses these oscillations more efficiently
and yields systematically higher fidelities. In particular, it reaches
the high-fidelity regime, $\mathcal{F}>0.99$, for
$t_f/\tau \gtrsim 20$, while keeping the constraint error below
$\xi^2\leq 5\times10^{-2}$.}

\subsection{Sensitivity of fidelity}

We now examine the sensitivity of the final fidelity to finite relative errors in the nonlinear control parameters. Such errors may arise from imperfect calibration of the scattering lengths during Feshbach-resonance tuning. We introduce a relative control error $\varepsilon$ through
\begin{equation}
    \lambda^{(\varepsilon)}(t)
    =
    \left(1+\varepsilon\right)\lambda(t),
\end{equation}
where $\lambda(t)$ denotes the nonlinear strength being tested. In Case II, $\lambda(t)=\delta g(t)$ is the controlled MF nonlinearity,
whereas in Case III, $\lambda(t)=g(t)$ is the controlled BMF
nonlinearity.

For each value of $\varepsilon$, the system is propagated with the perturbed protocol $\lambda^{(\varepsilon)}(t)$, and the final fidelity is evaluated as
\begin{equation}
\mathcal{F}(t_f,\varepsilon)
=
\frac{
\left|
\int dx\,\Psi_f^*(x)\psi^{(\varepsilon)}(x,t_f)
\right|^2
}
{
\left(\int dx\,|\Psi_f(x)|^2\right)
\left(\int dx\,|\psi^{(\varepsilon)}(x,t_f)|^2\right)
}.
\label{eq:fidelity_error}
\end{equation}
Instead of using a local derivative at $\varepsilon=0$, we quantify the response over a finite interval of relative errors. We define the sensitivity
\begin{equation}
    S(t_f;\varepsilon_0)
    =
    \frac{1}{2\varepsilon_0}
    \int_{-\varepsilon_0}^{\varepsilon_0}
    \left[1-\mathcal{F}(t_f,\varepsilon)\right]
    d\varepsilon .
    \label{eq:sensitivity}
\end{equation}
This quantity is the average final infidelity under bounded relative errors $|\varepsilon|\leq\varepsilon_0$. A smaller value of $S$ therefore indicates that the protocol remains closer to the target state in the presence of finite control fluctuations.

\begin{figure}[t]
\begin{minipage}{\linewidth}
  \centering
  \makebox[0pt][l]{\raisebox{9ex}{\hspace{13mm} \large(a)}}
\includegraphics[width=0.86\columnwidth]{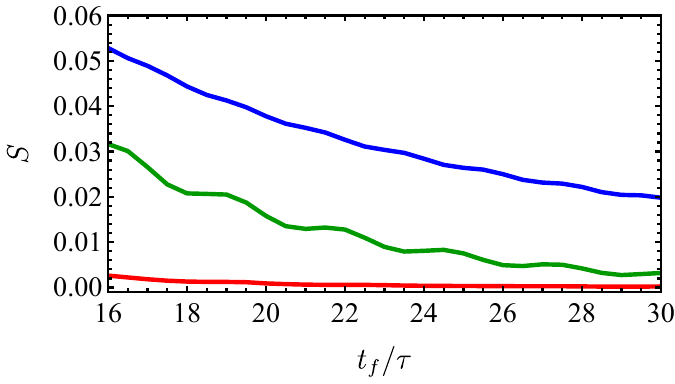}\\
\end{minipage}
\begin{minipage}{\linewidth}
  \centering
  \makebox[0pt][l]{\raisebox{8ex}{\hspace{13mm} \large(b)}}
\includegraphics[width=0.86\columnwidth]{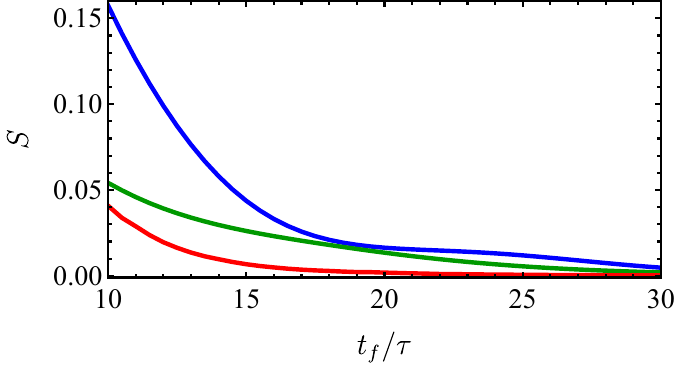}
\end{minipage}
\caption{
Sensitivity to finite relative errors in the nonlinear control
parameter. The quantity $S(t_f;\varepsilon_0)$ is evaluated with
$\varepsilon_0=0.004$. Panels (a) and (b) correspond to the constrained
cases with fixed BMF nonlinearity and fixed MF nonlinearity,
respectively. The blue, green, and red curves show
the reference, forced STA, and optimised STA protocols.
}
\label{fig:sensitivity}
\end{figure}

{We concentrate here on the cases II and III with constraints.}
Figure~\ref{fig:sensitivity} shows $S$ as a function of the final time for the reference, forced STA, and optimised STA protocols. In both constrained scenarios, the optimised STA protocol gives the smallest average final infidelity
over the tested error interval. The improvement is particularly pronounced at short final times, where the reference and forced protocols retain larger finite-time errors. As $t_f$ increases, the sensitivity decreases for all protocols, consistent with the approach toward adiabatic following under slower driving.

\subsection{Experimental relevance}

We briefly comment on the experimental relevance of the parameter
regime considered here. Quantum droplets have been realized in
ultracold Bose mixtures with tunable interactions, where Feshbach
resonances provide a direct handle for adjusting the scattering
lengths in time
\cite{Tarruellscience,Fattori2018,Tarruell2018PRL,Burchianti2025}.
Recent studies have further extended quantum-droplet physics to
different geometries and low-dimensional mixtures
\cite{Li2025SphericalDroplets,Xiao2026Asymmetric1D,he2025}. In such systems, the effective MF and
BMF nonlinearities are determined by the intra- and interspecies
interaction strengths. Therefore, the designed functions $g(t)$ and
$\delta g(t)$ can be regarded as target interaction trajectories for
fast dynamical control. For typical elongated ultracold-gas
experiments, the transverse oscillator length and time scales are on
the order of micrometers and milliseconds, respectively. Thus, the
dimensionless final times used in our simulations correspond to
experimentally relevant dynamical time scales. The explicit conversion
from the target nonlinear trajectories to a magnetic-field ramp is
system dependent and can be incorporated once a particular atomic
mixture and Feshbach resonance are specified.


\section{Conclusion}
\label{sec:6}

We have developed a variational inverse-engineering approach for the fast dynamical control of one-dimensional quantum droplets. By tuning the effective MF and BMF nonlinearities, which can be related to the intra- and interspecies scattering lengths of a binary mixture, the method connects prescribed initial and target self-bound states within short final times. The use of a bright-droplet variational ansatz allows the protocol to capture both narrow soliton-like configurations and broader flat-top droplets.

We first considered the unconstrained case, in which both nonlinearities are allowed to vary in time. We then analyzed two constrained scenarios: fixed BMF nonlinearity with time-dependent MF interaction, and fixed MF nonlinearity with time-dependent BMF interaction. In the constrained cases, additional free parameters in the auxiliary trajectories provide sufficient flexibility to reduce the constraint error and substantially improve the final fidelity. Direct simulations of the extended Gross-Pitaevskii equation confirm that optimised STA protocols can reach fidelities $\mathcal{F}>0.99$ on time scales much shorter than those required by simple reference ramps.

We also quantified the sensitivity of the final fidelity to finite relative errors in the nonlinear control strength. The optimised protocols exhibit the lowest average infidelity over the tested error interval, showing that the optimisation improves not only the speed of the transfer but also its robustness. These results demonstrate that inverse-engineered interaction protocols provide a useful tool for controlled state preparation and manipulation in self-bound quantum fluids.

We have considered two example cases of constraints, however we want to underline that the same strategy can be extended to other
types of constraints.

Future work could use this framework to identify the minimum ramp time
for a coherent soliton--droplet conversion under interaction tuning
\cite{Tarruell2018PRL}, and to determine whether optimised nonlinear
ramps can suppress the dominant breathing-mode excitation in
one-dimensional asymmetric droplets \cite{Xiao2026Asymmetric1D}.

\begin{acknowledgements}
Special thanks to D. Wong, R. Wang, J. Zhang and X. Chen for their insightful discussions at the early stages of this project. We are grateful to D. Rea for the fruitful discussion and careful reading of the manuscript. 
J.L. acknowledges that this publication was partly supported by Research Ireland under Grant No. 24/PATH-S/12716.
A.R. acknowledges that this publication has emanated from research
conducted with the support of Taighde Éireann – Research Ireland, under Grant number 19/FFP/6951 and under Grant number 23/RC/12197 at Rinn Quantum.
This work was supported by the Okinawa Institute of Science and Technology Graduate School (OIST) and
the JST Grant No. JPMJPF2221. T.F. acknowledges support from JSPS KAKENHI Grant No. JP23K03290.
\end{acknowledgements}
%


\begin{thebibliography}{37}%
\makeatletter
\providecommand \@ifxundefined [1]{%
 \@ifx{#1\undefined}
}%
\providecommand \@ifnum [1]{%
 \ifnum #1\expandafter \@firstoftwo
 \else \expandafter \@secondoftwo
 \fi
}%
\providecommand \@ifx [1]{%
 \ifx #1\expandafter \@firstoftwo
 \else \expandafter \@secondoftwo
 \fi
}%
\providecommand \natexlab [1]{#1}%
\providecommand \enquote  [1]{``#1''}%
\providecommand \bibnamefont  [1]{#1}%
\providecommand \bibfnamefont [1]{#1}%
\providecommand \citenamefont [1]{#1}%
\providecommand \href@noop [0]{\@secondoftwo}%
\providecommand \href [0]{\begingroup \@sanitize@url \@href}%
\providecommand \@href[1]{\@@startlink{#1}\@@href}%
\providecommand \@@href[1]{\endgroup#1\@@endlink}%
\providecommand \@sanitize@url [0]{\catcode `\\12\catcode `\$12\catcode
  `\&12\catcode `\#12\catcode `\^12\catcode `\_12\catcode `\%12\relax}%
\providecommand \@@startlink[1]{}%
\providecommand \@@endlink[0]{}%
\providecommand \url  [0]{\begingroup\@sanitize@url \@url }%
\providecommand \@url [1]{\endgroup\@href {#1}{\urlprefix }}%
\providecommand \urlprefix  [0]{URL }%
\providecommand \Eprint [0]{\href }%
\providecommand \doibase [0]{https://doi.org/}%
\providecommand \selectlanguage [0]{\@gobble}%
\providecommand \bibinfo  [0]{\@secondoftwo}%
\providecommand \bibfield  [0]{\@secondoftwo}%
\providecommand \translation [1]{[#1]}%
\providecommand \BibitemOpen [0]{}%
\providecommand \bibitemStop [0]{}%
\providecommand \bibitemNoStop [0]{.\EOS\space}%
\providecommand \EOS [0]{\spacefactor3000\relax}%
\providecommand \BibitemShut  [1]{\csname bibitem#1\endcsname}%
\let\auto@bib@innerbib\@empty
\bibitem [{\citenamefont {Cabrera}\ \emph {et~al.}(2018)\citenamefont
  {Cabrera}, \citenamefont {Tanzi}, \citenamefont {Sanz}, \citenamefont
  {Naylor}, \citenamefont {Thomas}, \citenamefont {Cheiney},\ and\
  \citenamefont {Tarruell}}]{Tarruellscience}%
  \BibitemOpen
  \bibfield  {author} {\bibinfo {author} {\bibfnamefont {C.~R.}\ \bibnamefont
  {Cabrera}}, \bibinfo {author} {\bibfnamefont {L.}~\bibnamefont {Tanzi}},
  \bibinfo {author} {\bibfnamefont {J.}~\bibnamefont {Sanz}}, \bibinfo {author}
  {\bibfnamefont {B.}~\bibnamefont {Naylor}}, \bibinfo {author} {\bibfnamefont
  {P.}~\bibnamefont {Thomas}}, \bibinfo {author} {\bibfnamefont
  {P.}~\bibnamefont {Cheiney}},\ and\ \bibinfo {author} {\bibfnamefont
  {L.}~\bibnamefont {Tarruell}},\ }\bibfield  {title} {\bibinfo {title}
  {Quantum liquid droplets in a mixture of bose-einstein condensates},\ }\href
  {https://doi.org/10.1126/science.aao5686} {\bibfield  {journal} {\bibinfo
  {journal} {Science}\ }\textbf {\bibinfo {volume} {359}},\ \bibinfo {pages}
  {301} (\bibinfo {year} {2018})}\BibitemShut {NoStop}%
\bibitem [{\citenamefont {Ferrier-Barbut}\ \emph {et~al.}(2016)\citenamefont
  {Ferrier-Barbut}, \citenamefont {Kadau}, \citenamefont {Schmitt},
  \citenamefont {Wenzel},\ and\ \citenamefont {Pfau}}]{Pfau2016}%
  \BibitemOpen
  \bibfield  {author} {\bibinfo {author} {\bibfnamefont {I.}~\bibnamefont
  {Ferrier-Barbut}}, \bibinfo {author} {\bibfnamefont {H.}~\bibnamefont
  {Kadau}}, \bibinfo {author} {\bibfnamefont {M.}~\bibnamefont {Schmitt}},
  \bibinfo {author} {\bibfnamefont {M.}~\bibnamefont {Wenzel}},\ and\ \bibinfo
  {author} {\bibfnamefont {T.}~\bibnamefont {Pfau}},\ }\bibfield  {title}
  {\bibinfo {title} {Observation of quantum droplets in a strongly dipolar bose
  gas},\ }\href {https://doi.org/10.1103/PhysRevLett.116.215301} {\bibfield
  {journal} {\bibinfo  {journal} {Phys. Rev. Lett.}\ }\textbf {\bibinfo
  {volume} {116}},\ \bibinfo {pages} {215301} (\bibinfo {year}
  {2016})}\BibitemShut {NoStop}%
\bibitem [{\citenamefont {Semeghini}\ \emph {et~al.}(2018)\citenamefont
  {Semeghini}, \citenamefont {Ferioli}, \citenamefont {Masi}, \citenamefont
  {Mazzinghi}, \citenamefont {Wolswijk}, \citenamefont {Minardi}, \citenamefont
  {Modugno}, \citenamefont {Modugno}, \citenamefont {Inguscio},\ and\
  \citenamefont {Fattori}}]{Fattori2018}%
  \BibitemOpen
  \bibfield  {author} {\bibinfo {author} {\bibfnamefont {G.}~\bibnamefont
  {Semeghini}}, \bibinfo {author} {\bibfnamefont {G.}~\bibnamefont {Ferioli}},
  \bibinfo {author} {\bibfnamefont {L.}~\bibnamefont {Masi}}, \bibinfo {author}
  {\bibfnamefont {C.}~\bibnamefont {Mazzinghi}}, \bibinfo {author}
  {\bibfnamefont {L.}~\bibnamefont {Wolswijk}}, \bibinfo {author}
  {\bibfnamefont {F.}~\bibnamefont {Minardi}}, \bibinfo {author} {\bibfnamefont
  {M.}~\bibnamefont {Modugno}}, \bibinfo {author} {\bibfnamefont
  {G.}~\bibnamefont {Modugno}}, \bibinfo {author} {\bibfnamefont
  {M.}~\bibnamefont {Inguscio}},\ and\ \bibinfo {author} {\bibfnamefont
  {M.}~\bibnamefont {Fattori}},\ }\bibfield  {title} {\bibinfo {title}
  {Self-bound quantum droplets of atomic mixtures in free space},\ }\href
  {https://doi.org/10.1103/PhysRevLett.120.235301} {\bibfield  {journal}
  {\bibinfo  {journal} {Phys. Rev. Lett.}\ }\textbf {\bibinfo {volume} {120}},\
  \bibinfo {pages} {235301} (\bibinfo {year} {2018})}\BibitemShut {NoStop}%
\bibitem [{\citenamefont {Chomaz}\ \emph {et~al.}(2022)\citenamefont {Chomaz},
  \citenamefont {Ferrier-Barbut}, \citenamefont {Ferlaino}, \citenamefont
  {Laburthe-Tolra}, \citenamefont {Lev},\ and\ \citenamefont
  {Pfau}}]{Chomaz_2023}%
  \BibitemOpen
  \bibfield  {author} {\bibinfo {author} {\bibfnamefont {L.}~\bibnamefont
  {Chomaz}}, \bibinfo {author} {\bibfnamefont {I.}~\bibnamefont
  {Ferrier-Barbut}}, \bibinfo {author} {\bibfnamefont {F.}~\bibnamefont
  {Ferlaino}}, \bibinfo {author} {\bibfnamefont {B.}~\bibnamefont
  {Laburthe-Tolra}}, \bibinfo {author} {\bibfnamefont {B.~L.}\ \bibnamefont
  {Lev}},\ and\ \bibinfo {author} {\bibfnamefont {T.}~\bibnamefont {Pfau}},\
  }\bibfield  {title} {\bibinfo {title} {Dipolar physics: a review of
  experiments with magnetic quantum gases},\ }\href
  {https://doi.org/10.1088/1361-6633/aca814} {\bibfield  {journal} {\bibinfo
  {journal} {Reports on Progress in Physics}\ }\textbf {\bibinfo {volume}
  {86}},\ \bibinfo {pages} {026401} (\bibinfo {year} {2022})}\BibitemShut
  {NoStop}%
\bibitem [{\citenamefont {Petrov}(2015)}]{Petrov}%
  \BibitemOpen
  \bibfield  {author} {\bibinfo {author} {\bibfnamefont {D.~S.}\ \bibnamefont
  {Petrov}},\ }\bibfield  {title} {\bibinfo {title} {Quantum mechanical
  stabilization of a collapsing bose-bose mixture},\ }\href
  {https://doi.org/10.1103/PhysRevLett.115.155302} {\bibfield  {journal}
  {\bibinfo  {journal} {Physical Review Letters}\ }\textbf {\bibinfo {volume}
  {115}},\ \bibinfo {pages} {155302} (\bibinfo {year} {2015})}\BibitemShut
  {NoStop}%
\bibitem [{\citenamefont {Astrakharchik}\ and\ \citenamefont
  {Malomed}(2018)}]{Malomed2018}%
  \BibitemOpen
  \bibfield  {author} {\bibinfo {author} {\bibfnamefont {G.~E.}\ \bibnamefont
  {Astrakharchik}}\ and\ \bibinfo {author} {\bibfnamefont {B.~A.}\ \bibnamefont
  {Malomed}},\ }\bibfield  {title} {\bibinfo {title} {Dynamics of
  one-dimensional quantum droplets},\ }\href
  {https://doi.org/10.1103/PhysRevA.98.013631} {\bibfield  {journal} {\bibinfo
  {journal} {Phys. Rev. A}\ }\textbf {\bibinfo {volume} {98}},\ \bibinfo
  {pages} {013631} (\bibinfo {year} {2018})}\BibitemShut {NoStop}%
\bibitem [{\citenamefont {Lee}\ \emph {et~al.}(1957)\citenamefont {Lee},
  \citenamefont {Huang},\ and\ \citenamefont {Yang}}]{LHY1957}%
  \BibitemOpen
  \bibfield  {author} {\bibinfo {author} {\bibfnamefont {T.~D.}\ \bibnamefont
  {Lee}}, \bibinfo {author} {\bibfnamefont {K.}~\bibnamefont {Huang}},\ and\
  \bibinfo {author} {\bibfnamefont {C.~N.}\ \bibnamefont {Yang}},\ }\bibfield
  {title} {\bibinfo {title} {Eigenvalues and eigenfunctions of a bose system of
  hard spheres and its low-temperature properties},\ }\href
  {https://doi.org/10.1103/PhysRev.106.1135} {\bibfield  {journal} {\bibinfo
  {journal} {Phys. Rev.}\ }\textbf {\bibinfo {volume} {106}},\ \bibinfo {pages}
  {1135} (\bibinfo {year} {1957})}\BibitemShut {NoStop}%
\bibitem [{\citenamefont {Edmonds}(2023)}]{Matthew22}%
  \BibitemOpen
  \bibfield  {author} {\bibinfo {author} {\bibfnamefont {M.}~\bibnamefont
  {Edmonds}},\ }\bibfield  {title} {\bibinfo {title} {Dark quantum droplets and
  solitary waves in beyond-mean-field bose-einstein condensate mixtures},\
  }\href {https://doi.org/10.1103/PhysRevResearch.5.023175} {\bibfield
  {journal} {\bibinfo  {journal} {Phys. Rev. Res.}\ }\textbf {\bibinfo {volume}
  {5}},\ \bibinfo {pages} {023175} (\bibinfo {year} {2023})}\BibitemShut
  {NoStop}%
\bibitem [{\citenamefont {Katsimiga}\ \emph {et~al.}(2023)\citenamefont
  {Katsimiga}, \citenamefont {Mistakidis}, \citenamefont {Koutsokostas},
  \citenamefont {Frantzeskakis}, \citenamefont {Carretero-Gonz\'alez},\ and\
  \citenamefont {Kevrekidis}}]{Kevrekidis23}%
  \BibitemOpen
  \bibfield  {author} {\bibinfo {author} {\bibfnamefont {G.~C.}\ \bibnamefont
  {Katsimiga}}, \bibinfo {author} {\bibfnamefont {S.~I.}\ \bibnamefont
  {Mistakidis}}, \bibinfo {author} {\bibfnamefont {G.~N.}\ \bibnamefont
  {Koutsokostas}}, \bibinfo {author} {\bibfnamefont {D.~J.}\ \bibnamefont
  {Frantzeskakis}}, \bibinfo {author} {\bibfnamefont {R.}~\bibnamefont
  {Carretero-Gonz\'alez}},\ and\ \bibinfo {author} {\bibfnamefont {P.~G.}\
  \bibnamefont {Kevrekidis}},\ }\bibfield  {title} {\bibinfo {title} {Solitary
  waves in a quantum droplet-bearing system},\ }\href
  {https://doi.org/10.1103/PhysRevA.107.063308} {\bibfield  {journal} {\bibinfo
   {journal} {Phys. Rev. A}\ }\textbf {\bibinfo {volume} {107}},\ \bibinfo
  {pages} {063308} (\bibinfo {year} {2023})}\BibitemShut {NoStop}%
\bibitem [{\citenamefont {Tylutki}\ \emph {et~al.}(2020)\citenamefont
  {Tylutki}, \citenamefont {Astrakharchik}, \citenamefont {Malomed},\ and\
  \citenamefont {Petrov}}]{Dmitry2020}%
  \BibitemOpen
  \bibfield  {author} {\bibinfo {author} {\bibfnamefont {M.}~\bibnamefont
  {Tylutki}}, \bibinfo {author} {\bibfnamefont {G.~E.}\ \bibnamefont
  {Astrakharchik}}, \bibinfo {author} {\bibfnamefont {B.~A.}\ \bibnamefont
  {Malomed}},\ and\ \bibinfo {author} {\bibfnamefont {D.~S.}\ \bibnamefont
  {Petrov}},\ }\bibfield  {title} {\bibinfo {title} {Collective excitations of
  a one-dimensional quantum droplet},\ }\href
  {https://doi.org/10.1103/PhysRevA.101.051601} {\bibfield  {journal} {\bibinfo
   {journal} {Phys. Rev. A}\ }\textbf {\bibinfo {volume} {101}},\ \bibinfo
  {pages} {051601} (\bibinfo {year} {2020})}\BibitemShut {NoStop}%
\bibitem [{\citenamefont {Halder}\ \emph {et~al.}(2022)\citenamefont {Halder},
  \citenamefont {Mukherjee}, \citenamefont {Mistakidis}, \citenamefont {Das},
  \citenamefont {Kevrekidis}, \citenamefont {Panigrahi}, \citenamefont
  {Majumder},\ and\ \citenamefont {Sadeghpour}}]{Sadeghpour2022}%
  \BibitemOpen
  \bibfield  {author} {\bibinfo {author} {\bibfnamefont {S.}~\bibnamefont
  {Halder}}, \bibinfo {author} {\bibfnamefont {K.}~\bibnamefont {Mukherjee}},
  \bibinfo {author} {\bibfnamefont {S.~I.}\ \bibnamefont {Mistakidis}},
  \bibinfo {author} {\bibfnamefont {S.}~\bibnamefont {Das}}, \bibinfo {author}
  {\bibfnamefont {P.~G.}\ \bibnamefont {Kevrekidis}}, \bibinfo {author}
  {\bibfnamefont {P.~K.}\ \bibnamefont {Panigrahi}}, \bibinfo {author}
  {\bibfnamefont {S.}~\bibnamefont {Majumder}},\ and\ \bibinfo {author}
  {\bibfnamefont {H.~R.}\ \bibnamefont {Sadeghpour}},\ }\bibfield  {title}
  {\bibinfo {title} {Phase diagram and dynamics of $^{164}\mathrm{Dy}$ dipolar
  bose-einstein condensate in the presence of a rotating anisotropic magnetic
  field},\ }\href {https://doi.org/https://doi.org/10.48550/arXiv.2205.05193}
  {\bibfield  {journal} {\bibinfo  {journal} {arXiv}\ ,\ \bibinfo {pages}
  {2205.05193}} (\bibinfo {year} {2022})}\BibitemShut {NoStop}%
\bibitem [{\citenamefont {Bland}\ \emph {et~al.}(2022)\citenamefont {Bland},
  \citenamefont {Poli}, \citenamefont {Politi}, \citenamefont {Klaus},
  \citenamefont {Norcia}, \citenamefont {Ferlaino}, \citenamefont {Santos},\
  and\ \citenamefont {Bisset}}]{Bisset2022}%
  \BibitemOpen
  \bibfield  {author} {\bibinfo {author} {\bibfnamefont {T.}~\bibnamefont
  {Bland}}, \bibinfo {author} {\bibfnamefont {E.}~\bibnamefont {Poli}},
  \bibinfo {author} {\bibfnamefont {C.}~\bibnamefont {Politi}}, \bibinfo
  {author} {\bibfnamefont {L.}~\bibnamefont {Klaus}}, \bibinfo {author}
  {\bibfnamefont {M.~A.}\ \bibnamefont {Norcia}}, \bibinfo {author}
  {\bibfnamefont {F.}~\bibnamefont {Ferlaino}}, \bibinfo {author}
  {\bibfnamefont {L.}~\bibnamefont {Santos}},\ and\ \bibinfo {author}
  {\bibfnamefont {R.~N.}\ \bibnamefont {Bisset}},\ }\bibfield  {title}
  {\bibinfo {title} {Two-dimensional supersolid formation in dipolar
  condensates},\ }\href {https://doi.org/10.1103/PhysRevLett.128.195302}
  {\bibfield  {journal} {\bibinfo  {journal} {Phys. Rev. Lett.}\ }\textbf
  {\bibinfo {volume} {128}},\ \bibinfo {pages} {195302} (\bibinfo {year}
  {2022})}\BibitemShut {NoStop}%
\bibitem [{\citenamefont {Nie}\ \emph {et~al.}(2023)\citenamefont {Nie},
  \citenamefont {Zheng},\ and\ \citenamefont {Yang}}]{Yang2023}%
  \BibitemOpen
  \bibfield  {author} {\bibinfo {author} {\bibfnamefont {Y.}~\bibnamefont
  {Nie}}, \bibinfo {author} {\bibfnamefont {J.-H.}\ \bibnamefont {Zheng}},\
  and\ \bibinfo {author} {\bibfnamefont {T.}~\bibnamefont {Yang}},\ }\bibfield
  {title} {\bibinfo {title} {Spectra and dynamics of quantum droplets in an
  optical lattice},\ }\href {https://doi.org/10.1103/PhysRevA.108.053310}
  {\bibfield  {journal} {\bibinfo  {journal} {Phys. Rev. A}\ }\textbf {\bibinfo
  {volume} {108}},\ \bibinfo {pages} {053310} (\bibinfo {year}
  {2023})}\BibitemShut {NoStop}%
\bibitem [{\citenamefont {Ferioli}\ \emph {et~al.}(2019)\citenamefont
  {Ferioli}, \citenamefont {Semeghini}, \citenamefont {Masi}, \citenamefont
  {Giusti}, \citenamefont {Modugno}, \citenamefont {Inguscio}, \citenamefont
  {Gallem\'{\i}}, \citenamefont {Recati},\ and\ \citenamefont
  {Fattori}}]{Fattori2019}%
  \BibitemOpen
  \bibfield  {author} {\bibinfo {author} {\bibfnamefont {G.}~\bibnamefont
  {Ferioli}}, \bibinfo {author} {\bibfnamefont {G.}~\bibnamefont {Semeghini}},
  \bibinfo {author} {\bibfnamefont {L.}~\bibnamefont {Masi}}, \bibinfo {author}
  {\bibfnamefont {G.}~\bibnamefont {Giusti}}, \bibinfo {author} {\bibfnamefont
  {G.}~\bibnamefont {Modugno}}, \bibinfo {author} {\bibfnamefont
  {M.}~\bibnamefont {Inguscio}}, \bibinfo {author} {\bibfnamefont
  {A.}~\bibnamefont {Gallem\'{\i}}}, \bibinfo {author} {\bibfnamefont
  {A.}~\bibnamefont {Recati}},\ and\ \bibinfo {author} {\bibfnamefont
  {M.}~\bibnamefont {Fattori}},\ }\bibfield  {title} {\bibinfo {title}
  {Collisions of self-bound quantum droplets},\ }\href
  {https://doi.org/10.1103/PhysRevLett.122.090401} {\bibfield  {journal}
  {\bibinfo  {journal} {Phys. Rev. Lett.}\ }\textbf {\bibinfo {volume} {122}},\
  \bibinfo {pages} {090401} (\bibinfo {year} {2019})}\BibitemShut {NoStop}%
\bibitem [{\citenamefont {Cikojevi{\'c}}\ \emph {et~al.}(2021)\citenamefont
  {Cikojevi{\'c}}, \citenamefont {Vranje{\v{s}}~Marki{\'c}}, \citenamefont
  {Pi}, \citenamefont {Barranco}, \citenamefont {Ancilotto},\ and\
  \citenamefont {Boronat}}]{Boronat2021}%
  \BibitemOpen
  \bibfield  {author} {\bibinfo {author} {\bibfnamefont {V.}~\bibnamefont
  {Cikojevi{\'c}}}, \bibinfo {author} {\bibfnamefont {L.}~\bibnamefont
  {Vranje{\v{s}}~Marki{\'c}}}, \bibinfo {author} {\bibfnamefont
  {M.}~\bibnamefont {Pi}}, \bibinfo {author} {\bibfnamefont {M.}~\bibnamefont
  {Barranco}}, \bibinfo {author} {\bibfnamefont {F.}~\bibnamefont
  {Ancilotto}},\ and\ \bibinfo {author} {\bibfnamefont {J.}~\bibnamefont
  {Boronat}},\ }\bibfield  {title} {\bibinfo {title} {Dynamics of equilibration
  and collisions in ultradilute quantum droplets},\ }\href
  {https://doi.org/10.1103/PhysRevResearch.3.043139} {\bibfield  {journal}
  {\bibinfo  {journal} {Phys. Rev. Research}\ }\textbf {\bibinfo {volume}
  {3}},\ \bibinfo {pages} {043139} (\bibinfo {year} {2021})}\BibitemShut
  {NoStop}%
\bibitem [{\citenamefont {Hu}\ \emph {et~al.}(2022)\citenamefont {Hu},
  \citenamefont {Fei}, \citenamefont {Chen},\ and\ \citenamefont
  {Zhang}}]{Hu2022}%
  \BibitemOpen
  \bibfield  {author} {\bibinfo {author} {\bibfnamefont {Y.}~\bibnamefont
  {Hu}}, \bibinfo {author} {\bibfnamefont {Y.}~\bibnamefont {Fei}}, \bibinfo
  {author} {\bibfnamefont {X.-L.}\ \bibnamefont {Chen}},\ and\ \bibinfo
  {author} {\bibfnamefont {Y.}~\bibnamefont {Zhang}},\ }\bibfield  {title}
  {\bibinfo {title} {Collisional dynamics of symmetric two-dimensional quantum
  droplets},\ }\href {https://doi.org/10.1007/s11467-022-1192-z} {\bibfield
  {journal} {\bibinfo  {journal} {Frontiers of Physics}\ }\textbf {\bibinfo
  {volume} {17}},\ \bibinfo {pages} {61505} (\bibinfo {year}
  {2022})}\BibitemShut {NoStop}%
\bibitem [{\citenamefont {Cavicchioli}\ \emph {et~al.}(2025)\citenamefont
  {Cavicchioli}, \citenamefont {Fort}, \citenamefont {Ancilotto}, \citenamefont
  {Modugno}, \citenamefont {Minardi},\ and\ \citenamefont
  {Burchianti}}]{Burchianti2025}%
  \BibitemOpen
  \bibfield  {author} {\bibinfo {author} {\bibfnamefont {L.}~\bibnamefont
  {Cavicchioli}}, \bibinfo {author} {\bibfnamefont {C.}~\bibnamefont {Fort}},
  \bibinfo {author} {\bibfnamefont {F.}~\bibnamefont {Ancilotto}}, \bibinfo
  {author} {\bibfnamefont {M.}~\bibnamefont {Modugno}}, \bibinfo {author}
  {\bibfnamefont {F.}~\bibnamefont {Minardi}},\ and\ \bibinfo {author}
  {\bibfnamefont {A.}~\bibnamefont {Burchianti}},\ }\bibfield  {title}
  {\bibinfo {title} {Dynamical formation of multiple quantum droplets in a
  bose-bose mixture},\ }\href {https://doi.org/10.1103/PhysRevLett.134.093401}
  {\bibfield  {journal} {\bibinfo  {journal} {Phys. Rev. Lett.}\ }\textbf
  {\bibinfo {volume} {134}},\ \bibinfo {pages} {093401} (\bibinfo {year}
  {2025})}\BibitemShut {NoStop}%
\bibitem [{\citenamefont {Li}\ \emph {et~al.}(2025)\citenamefont {Li},
  \citenamefont {Gong}, \citenamefont {Zhou},\ and\ \citenamefont
  {Zhou}}]{Li2025SphericalDroplets}%
  \BibitemOpen
  \bibfield  {author} {\bibinfo {author} {\bibfnamefont {W.-H.}\ \bibnamefont
  {Li}}, \bibinfo {author} {\bibfnamefont {M.}~\bibnamefont {Gong}}, \bibinfo
  {author} {\bibfnamefont {Z.-W.}\ \bibnamefont {Zhou}},\ and\ \bibinfo
  {author} {\bibfnamefont {X.-F.}\ \bibnamefont {Zhou}},\ }\bibfield  {title}
  {\bibinfo {title} {Stabilization of bose-bose mixtures on a spherical surface
  induced by the lee-huang-yang correction},\ }\href
  {https://doi.org/10.1103/PhysRevA.111.043301} {\bibfield  {journal} {\bibinfo
   {journal} {Physical Review A}\ }\textbf {\bibinfo {volume} {111}},\ \bibinfo
  {pages} {043301} (\bibinfo {year} {2025})}\BibitemShut {NoStop}%
\bibitem [{\citenamefont {Xiao}\ \emph {et~al.}(2026)\citenamefont {Xiao},
  \citenamefont {Zhang}, \citenamefont {Liu}, \citenamefont {Du}, \citenamefont
  {Chen},\ and\ \citenamefont {Zhang}}]{Xiao2026Asymmetric1D}%
  \BibitemOpen
  \bibfield  {author} {\bibinfo {author} {\bibfnamefont {H.}~\bibnamefont
  {Xiao}}, \bibinfo {author} {\bibfnamefont {X.}~\bibnamefont {Zhang}},
  \bibinfo {author} {\bibfnamefont {J.}~\bibnamefont {Liu}}, \bibinfo {author}
  {\bibfnamefont {X.}~\bibnamefont {Du}}, \bibinfo {author} {\bibfnamefont
  {X.-L.}\ \bibnamefont {Chen}},\ and\ \bibinfo {author} {\bibfnamefont
  {Y.}~\bibnamefont {Zhang}},\ }\bibfield  {title} {\bibinfo {title}
  {One-dimensional asymmetrically interacting quantum droplets in bose-bose
  mixtures},\ }\href {https://doi.org/10.1103/2zbp-j5bg} {\bibfield  {journal}
  {\bibinfo  {journal} {Physical Review A}\ }\textbf {\bibinfo {volume}
  {113}},\ \bibinfo {pages} {063301} (\bibinfo {year} {2026})}\BibitemShut
  {NoStop}%
\bibitem [{\citenamefont {Gu\'ery-Odelin}\ \emph {et~al.}(2019)\citenamefont
  {Gu\'ery-Odelin}, \citenamefont {Ruschhaupt}, \citenamefont {Kiely},
  \citenamefont {Torrontegui}, \citenamefont {Mart\'{\i}nez-Garaot},\ and\
  \citenamefont {Muga}}]{ReviewSTA}%
  \BibitemOpen
  \bibfield  {author} {\bibinfo {author} {\bibfnamefont {D.}~\bibnamefont
  {Gu\'ery-Odelin}}, \bibinfo {author} {\bibfnamefont {A.}~\bibnamefont
  {Ruschhaupt}}, \bibinfo {author} {\bibfnamefont {A.}~\bibnamefont {Kiely}},
  \bibinfo {author} {\bibfnamefont {E.}~\bibnamefont {Torrontegui}}, \bibinfo
  {author} {\bibfnamefont {S.}~\bibnamefont {Mart\'{\i}nez-Garaot}},\ and\
  \bibinfo {author} {\bibfnamefont {J.~G.}\ \bibnamefont {Muga}},\ }\bibfield
  {title} {\bibinfo {title} {Shortcuts to adiabaticity: Concepts, methods, and
  applications},\ }\href {https://doi.org/10.1103/RevModPhys.91.045001}
  {\bibfield  {journal} {\bibinfo  {journal} {Rev. Mod. Phys.}\ }\textbf
  {\bibinfo {volume} {91}},\ \bibinfo {pages} {045001} (\bibinfo {year}
  {2019})}\BibitemShut {NoStop}%
\bibitem [{\citenamefont {Li}\ \emph {et~al.}(2016)\citenamefont {Li},
  \citenamefont {Sun},\ and\ \citenamefont {Chen}}]{Li2016}%
  \BibitemOpen
  \bibfield  {author} {\bibinfo {author} {\bibfnamefont {J.}~\bibnamefont
  {Li}}, \bibinfo {author} {\bibfnamefont {K.}~\bibnamefont {Sun}},\ and\
  \bibinfo {author} {\bibfnamefont {X.}~\bibnamefont {Chen}},\ }\bibfield
  {title} {\bibinfo {title} {Shortcut to adiabatic control of soliton matter
  waves by tunable interaction},\ }\href {https://doi.org/10.1038/srep38258}
  {\bibfield  {journal} {\bibinfo  {journal} {Scientific Reports}\ }\textbf
  {\bibinfo {volume} {6}},\ \bibinfo {pages} {38258} (\bibinfo {year}
  {2016})}\BibitemShut {NoStop}%
\bibitem [{\citenamefont {Fogarty}\ \emph {et~al.}(2019)\citenamefont
  {Fogarty}, \citenamefont {Ruks}, \citenamefont {Li},\ and\ \citenamefont
  {Busch}}]{Ruks:19}%
  \BibitemOpen
  \bibfield  {author} {\bibinfo {author} {\bibfnamefont {T.}~\bibnamefont
  {Fogarty}}, \bibinfo {author} {\bibfnamefont {L.}~\bibnamefont {Ruks}},
  \bibinfo {author} {\bibfnamefont {J.}~\bibnamefont {Li}},\ and\ \bibinfo
  {author} {\bibfnamefont {T.}~\bibnamefont {Busch}},\ }\bibfield  {title}
  {\bibinfo {title} {{Fast control of interactions in an ultracold two atom
  system: Managing correlations and irreversibility}},\ }\href
  {https://doi.org/10.21468/SciPostPhys.6.2.021} {\bibfield  {journal}
  {\bibinfo  {journal} {SciPost Phys.}\ }\textbf {\bibinfo {volume} {6}},\
  \bibinfo {pages} {021} (\bibinfo {year} {2019})}\BibitemShut {NoStop}%
\bibitem [{\citenamefont {Kahan}\ \emph {et~al.}(2019)\citenamefont {Kahan},
  \citenamefont {Fogarty}, \citenamefont {Li},\ and\ \citenamefont
  {Busch}}]{Kahan:19}%
  \BibitemOpen
  \bibfield  {author} {\bibinfo {author} {\bibfnamefont {A.}~\bibnamefont
  {Kahan}}, \bibinfo {author} {\bibfnamefont {T.}~\bibnamefont {Fogarty}},
  \bibinfo {author} {\bibfnamefont {J.}~\bibnamefont {Li}},\ and\ \bibinfo
  {author} {\bibfnamefont {T.}~\bibnamefont {Busch}},\ }\bibfield  {title}
  {\bibinfo {title} {{Driving Interactions Efficiently in a Composite Few-Body
  System}},\ }\href {https://doi.org/10.3390/universe5100207} {\bibfield
  {journal} {\bibinfo  {journal} {Universe}\ }\textbf {\bibinfo {volume} {5}},\
  \bibinfo {pages} {207} (\bibinfo {year} {2019})}\BibitemShut {NoStop}%
\bibitem [{\citenamefont {Hasan}\ \emph {et~al.}(2024)\citenamefont {Hasan},
  \citenamefont {Fogarty}, \citenamefont {Li}, \citenamefont {Ruschhaupt},\
  and\ \citenamefont {Busch}}]{Hasan2024}%
  \BibitemOpen
  \bibfield  {author} {\bibinfo {author} {\bibfnamefont {M.~S.}\ \bibnamefont
  {Hasan}}, \bibinfo {author} {\bibfnamefont {T.}~\bibnamefont {Fogarty}},
  \bibinfo {author} {\bibfnamefont {J.}~\bibnamefont {Li}}, \bibinfo {author}
  {\bibfnamefont {A.}~\bibnamefont {Ruschhaupt}},\ and\ \bibinfo {author}
  {\bibfnamefont {T.}~\bibnamefont {Busch}},\ }\bibfield  {title} {\bibinfo
  {title} {High fidelity control of a many-body tonks-girardeau gas with an
  effective mean-field approach},\ }\href
  {https://doi.org/10.1103/PhysRevResearch.6.023114} {\bibfield  {journal}
  {\bibinfo  {journal} {Phys. Rev. Res.}\ }\textbf {\bibinfo {volume} {6}},\
  \bibinfo {pages} {023114} (\bibinfo {year} {2024})}\BibitemShut {NoStop}%
\bibitem [{\citenamefont {Otajonov}\ \emph {et~al.}(2020)\citenamefont
  {Otajonov}, \citenamefont {Tsoy},\ and\ \citenamefont
  {Abdullaev}}]{Abdullaev20}%
  \BibitemOpen
  \bibfield  {author} {\bibinfo {author} {\bibfnamefont {S.~R.}\ \bibnamefont
  {Otajonov}}, \bibinfo {author} {\bibfnamefont {E.~N.}\ \bibnamefont {Tsoy}},\
  and\ \bibinfo {author} {\bibfnamefont {F.~K.}\ \bibnamefont {Abdullaev}},\
  }\bibfield  {title} {\bibinfo {title} {Variational approximation for
  two-dimensional quantum droplets},\ }\href
  {https://doi.org/10.1103/PhysRevE.102.062217} {\bibfield  {journal} {\bibinfo
   {journal} {Phys. Rev. E}\ }\textbf {\bibinfo {volume} {102}},\ \bibinfo
  {pages} {062217} (\bibinfo {year} {2020})}\BibitemShut {NoStop}%
\bibitem [{\citenamefont {Otajonov}\ \emph {et~al.}(2019)\citenamefont
  {Otajonov}, \citenamefont {Tsoy},\ and\ \citenamefont
  {Abdullaev}}]{Fatkhulla2019}%
  \BibitemOpen
  \bibfield  {author} {\bibinfo {author} {\bibfnamefont {S.~R.}\ \bibnamefont
  {Otajonov}}, \bibinfo {author} {\bibfnamefont {E.~N.}\ \bibnamefont {Tsoy}},\
  and\ \bibinfo {author} {\bibfnamefont {F.~K.}\ \bibnamefont {Abdullaev}},\
  }\bibfield  {title} {\bibinfo {title} {Stationary and dynamical properties of
  one-dimensional quantum droplets},\ }\href
  {https://doi.org/https://doi.org/10.1016/j.physleta.2019.125980} {\bibfield
  {journal} {\bibinfo  {journal} {Physics Letters A}\ }\textbf {\bibinfo
  {volume} {383}},\ \bibinfo {pages} {125980} (\bibinfo {year}
  {2019})}\BibitemShut {NoStop}%
\bibitem [{\citenamefont {St\"urmer}\ \emph {et~al.}(2021)\citenamefont
  {St\"urmer}, \citenamefont {Tengstrand}, \citenamefont {Sachdeva},\ and\
  \citenamefont {Reimann}}]{Reimann2021}%
  \BibitemOpen
  \bibfield  {author} {\bibinfo {author} {\bibfnamefont {P.}~\bibnamefont
  {St\"urmer}}, \bibinfo {author} {\bibfnamefont {M.~N.}\ \bibnamefont
  {Tengstrand}}, \bibinfo {author} {\bibfnamefont {R.}~\bibnamefont
  {Sachdeva}},\ and\ \bibinfo {author} {\bibfnamefont {S.~M.}\ \bibnamefont
  {Reimann}},\ }\bibfield  {title} {\bibinfo {title} {Breathing mode in
  two-dimensional binary self-bound bose-gas droplets},\ }\href
  {https://doi.org/10.1103/PhysRevA.103.053302} {\bibfield  {journal} {\bibinfo
   {journal} {Phys. Rev. A}\ }\textbf {\bibinfo {volume} {103}},\ \bibinfo
  {pages} {053302} (\bibinfo {year} {2021})}\BibitemShut {NoStop}%
\bibitem [{\citenamefont {Pelayo}\ \emph {et~al.}(2025)\citenamefont {Pelayo},
  \citenamefont {Bougas}, \citenamefont {Fogarty}, \citenamefont {Busch},\ and\
  \citenamefont {Mistakidis}}]{Mistakidis2025}%
  \BibitemOpen
  \bibfield  {author} {\bibinfo {author} {\bibfnamefont {J.~C.}\ \bibnamefont
  {Pelayo}}, \bibinfo {author} {\bibfnamefont {G.}~\bibnamefont {Bougas}},
  \bibinfo {author} {\bibfnamefont {T.}~\bibnamefont {Fogarty}}, \bibinfo
  {author} {\bibfnamefont {T.}~\bibnamefont {Busch}},\ and\ \bibinfo {author}
  {\bibfnamefont {S.~I.}\ \bibnamefont {Mistakidis}},\ }\bibfield  {title}
  {\bibinfo {title} {{Phases and dynamics of quantum droplets in the crossover
  to two-dimensions}},\ }\href {https://doi.org/10.21468/SciPostPhys.18.4.129}
  {\bibfield  {journal} {\bibinfo  {journal} {SciPost Phys.}\ }\textbf
  {\bibinfo {volume} {18}},\ \bibinfo {pages} {129} (\bibinfo {year}
  {2025})}\BibitemShut {NoStop}%
\bibitem [{\citenamefont {Otajonov}\ and\ \citenamefont
  {Abdullaev}(2026)}]{Fatkhulla2026}%
  \BibitemOpen
  \bibfield  {author} {\bibinfo {author} {\bibfnamefont {S.~R.}\ \bibnamefont
  {Otajonov}}\ and\ \bibinfo {author} {\bibfnamefont {F.~K.}\ \bibnamefont
  {Abdullaev}},\ }\bibfield  {title} {\bibinfo {title} {Josephson dynamics of
  two-dimensional bose-einstein condensates in a dual-core trap: Homogeneous,
  droplet-droplet, and vortex-vortex regimes},\ }\href
  {https://doi.org/10.1103/bphc-m85d} {\bibfield  {journal} {\bibinfo
  {journal} {Phys. Rev. A}\ }\textbf {\bibinfo {volume} {113}},\ \bibinfo
  {pages} {043319} (\bibinfo {year} {2026})}\BibitemShut {NoStop}%
\bibitem [{\citenamefont {Petrov}\ and\ \citenamefont
  {Astrakharchik}(2016)}]{LowD2016}%
  \BibitemOpen
  \bibfield  {author} {\bibinfo {author} {\bibfnamefont {D.~S.}\ \bibnamefont
  {Petrov}}\ and\ \bibinfo {author} {\bibfnamefont {G.~E.}\ \bibnamefont
  {Astrakharchik}},\ }\bibfield  {title} {\bibinfo {title} {Ultradilute
  low-dimensional liquids},\ }\href
  {https://doi.org/10.1103/PhysRevLett.117.100401} {\bibfield  {journal}
  {\bibinfo  {journal} {Phys. Rev. Lett.}\ }\textbf {\bibinfo {volume} {117}},\
  \bibinfo {pages} {100401} (\bibinfo {year} {2016})}\BibitemShut {NoStop}%
\bibitem [{\citenamefont {Pathak}\ and\ \citenamefont
  {Nath}(2022)}]{Pathak2022}%
  \BibitemOpen
  \bibfield  {author} {\bibinfo {author} {\bibfnamefont {M.~R.}\ \bibnamefont
  {Pathak}}\ and\ \bibinfo {author} {\bibfnamefont {A.}~\bibnamefont {Nath}},\
  }\bibfield  {title} {\bibinfo {title} {Dynamics of quantum droplets in an
  external harmonic confinement},\ }\href
  {https://doi.org/10.1038/s41598-022-10468-6} {\bibfield  {journal} {\bibinfo
  {journal} {Scientific Reports}\ }\textbf {\bibinfo {volume} {12}},\ \bibinfo
  {pages} {6904} (\bibinfo {year} {2022})}\BibitemShut {NoStop}%
\bibitem [{\citenamefont {Chin}\ \emph {et~al.}(2010)\citenamefont {Chin},
  \citenamefont {Grimm}, \citenamefont {Julienne},\ and\ \citenamefont
  {Tiesinga}}]{Feshbach}%
  \BibitemOpen
  \bibfield  {author} {\bibinfo {author} {\bibfnamefont {C.}~\bibnamefont
  {Chin}}, \bibinfo {author} {\bibfnamefont {R.}~\bibnamefont {Grimm}},
  \bibinfo {author} {\bibfnamefont {P.}~\bibnamefont {Julienne}},\ and\
  \bibinfo {author} {\bibfnamefont {E.}~\bibnamefont {Tiesinga}},\ }\bibfield
  {title} {\bibinfo {title} {Feshbach resonances in ultracold gases},\ }\href
  {https://doi.org/10.1103/RevModPhys.82.1225} {\bibfield  {journal} {\bibinfo
  {journal} {Rev. Mod. Phys.}\ }\textbf {\bibinfo {volume} {82}},\ \bibinfo
  {pages} {1225} (\bibinfo {year} {2010})}\BibitemShut {NoStop}%
\bibitem [{\citenamefont {P\'erez-Garc\'{\i}a}\ \emph
  {et~al.}(1997)\citenamefont {P\'erez-Garc\'{\i}a}, \citenamefont {Michinel},
  \citenamefont {Cirac}, \citenamefont {Lewenstein},\ and\ \citenamefont
  {Zoller}}]{Zoller1997}%
  \BibitemOpen
  \bibfield  {author} {\bibinfo {author} {\bibfnamefont {V.~M.}\ \bibnamefont
  {P\'erez-Garc\'{\i}a}}, \bibinfo {author} {\bibfnamefont {H.}~\bibnamefont
  {Michinel}}, \bibinfo {author} {\bibfnamefont {J.~I.}\ \bibnamefont {Cirac}},
  \bibinfo {author} {\bibfnamefont {M.}~\bibnamefont {Lewenstein}},\ and\
  \bibinfo {author} {\bibfnamefont {P.}~\bibnamefont {Zoller}},\ }\bibfield
  {title} {\bibinfo {title} {Dynamics of bose-einstein condensates: Variational
  solutions of the gross-pitaevskii equations},\ }\href
  {https://doi.org/10.1103/PhysRevA.56.1424} {\bibfield  {journal} {\bibinfo
  {journal} {Phys. Rev. A}\ }\textbf {\bibinfo {volume} {56}},\ \bibinfo
  {pages} {1424} (\bibinfo {year} {1997})}\BibitemShut {NoStop}%
\bibitem [{\citenamefont {Otajonov}\ \emph {et~al.}(2024)\citenamefont
  {Otajonov}, \citenamefont {Umarov},\ and\ \citenamefont
  {Abdullaev}}]{otajonov2024}%
  \BibitemOpen
  \bibfield  {author} {\bibinfo {author} {\bibfnamefont {S.~R.}\ \bibnamefont
  {Otajonov}}, \bibinfo {author} {\bibfnamefont {B.~A.}\ \bibnamefont
  {Umarov}},\ and\ \bibinfo {author} {\bibfnamefont {F.~K.}\ \bibnamefont
  {Abdullaev}},\ }\bibfield  {title} {\bibinfo {title} {Dynamics of
  quasi-one-dimensional quantum droplets in bose--bose mixtures},\ }\href@noop
  {} {\bibfield  {journal} {\bibinfo  {journal} {Chaos, Solitons \& Fractals}\
  }\textbf {\bibinfo {volume} {186}},\ \bibinfo {pages} {115212} (\bibinfo
  {year} {2024})}\BibitemShut {NoStop}%
\bibitem [{\citenamefont {Bao}\ and\ \citenamefont {Du}(2004)}]{BaoDu2004}%
  \BibitemOpen
  \bibfield  {author} {\bibinfo {author} {\bibfnamefont {W.}~\bibnamefont
  {Bao}}\ and\ \bibinfo {author} {\bibfnamefont {Q.}~\bibnamefont {Du}},\
  }\bibfield  {title} {\bibinfo {title} {Computing the ground state solution of
  {Bose--Einstein} condensates by a normalized gradient flow},\ }\href
  {https://doi.org/10.1137/S1064827503422956} {\bibfield  {journal} {\bibinfo
  {journal} {SIAM Journal on Scientific Computing}\ }\textbf {\bibinfo {volume}
  {25}},\ \bibinfo {pages} {1674} (\bibinfo {year} {2004})}\BibitemShut
  {NoStop}%
\bibitem [{\citenamefont {Cheiney}\ \emph {et~al.}(2018)\citenamefont
  {Cheiney}, \citenamefont {Cabrera}, \citenamefont {Sanz}, \citenamefont
  {Naylor}, \citenamefont {Tanzi},\ and\ \citenamefont
  {Tarruell}}]{Tarruell2018PRL}%
  \BibitemOpen
  \bibfield  {author} {\bibinfo {author} {\bibfnamefont {P.}~\bibnamefont
  {Cheiney}}, \bibinfo {author} {\bibfnamefont {C.~R.}\ \bibnamefont
  {Cabrera}}, \bibinfo {author} {\bibfnamefont {J.}~\bibnamefont {Sanz}},
  \bibinfo {author} {\bibfnamefont {B.}~\bibnamefont {Naylor}}, \bibinfo
  {author} {\bibfnamefont {L.}~\bibnamefont {Tanzi}},\ and\ \bibinfo {author}
  {\bibfnamefont {L.}~\bibnamefont {Tarruell}},\ }\bibfield  {title} {\bibinfo
  {title} {Bright soliton to quantum droplet transition in a mixture of
  bose-einstein condensates},\ }\href
  {https://doi.org/10.1103/PhysRevLett.120.135301} {\bibfield  {journal}
  {\bibinfo  {journal} {Phys. Rev. Lett.}\ }\textbf {\bibinfo {volume} {120}},\
  \bibinfo {pages} {135301} (\bibinfo {year} {2018})}\BibitemShut {NoStop}%
\bibitem [{\citenamefont {He}\ \emph {et~al.}(2025)\citenamefont {He},
  \citenamefont {Yi},\ and\ \citenamefont {Busch}}]{he2025}%
  \BibitemOpen
  \bibfield  {author} {\bibinfo {author} {\bibfnamefont {W.-B.}\ \bibnamefont
  {He}}, \bibinfo {author} {\bibfnamefont {S.}~\bibnamefont {Yi}},\ and\
  \bibinfo {author} {\bibfnamefont {T.}~\bibnamefont {Busch}},\ }\href
  {https://arxiv.org/abs/2508.16115} {\bibinfo {title} {Quantum droplets in
  one-dimensional mixtures of quasi bose-einstein condensates and
  tonks-girardeau gases}} (\bibinfo {year} {2025}),\ \Eprint
  {https://arxiv.org/abs/2508.16115} {arXiv:2508.16115 [cond-mat.quant-gas]}
  \BibitemShut {NoStop}%
\end{thebibliography}
\end{document}